\documentclass[11pt]{article}
\pdfoutput=1
\usepackage{jheppub}

\usepackage{lipsum}  
\usepackage{graphicx}
\usepackage{amsmath}
\usepackage{amssymb}
\usepackage{cancel}

\def\ba{\begin{eqnarray}}
\def\ea{\end{eqnarray}}

\def\be{\begin{equation}}
\def\ee{\end{equation}}

\DeclareFontFamily{OT1}{pzc}{}
\DeclareFontShape{OT1}{pzc}{m}{it}{ <-> s*[1.15] pzcmi7t }{}
\DeclareMathAlphabet{\mathpzc}{OT1}{pzc}{m}{it}

\newcommand{\wbar}{{\overline \tau}}
\newcommand{\Mbar}{{\overline{\mathcal M}}\hspace{0.5mm}}

\title{Far-from-equilibrium attractors for massive kinetic theory in the relaxation time approximation}

\author{H. Alalawi and M. Strickland}
\affiliation{Department of Physics, Kent State University, Kent, OH 44242 United States}

\emailAdd{halalawi@kent.edu}
\emailAdd{mstrick6@kent.edu}

\abstract{
We investigate whether early and late time attractors for non-conformal kinetic theories exist by computing the time-evolution of a large set of moments of the one-particle distribution function.  For this purpose we make use of a previously obtained exact solution of the 0+1D boost-invariant massive Boltzmann equation in relaxation time approximation.  We extend prior attractor studies of non-conformal systems by using a realistic mass- and temperature-dependent relaxation time and explicitly computing the effect of varying both the initial momentum-space anisotropy and initialization time on the time evolution of a large set of integral moments.  Our findings are consistent with prior studies, which found that there is an attractor for the scaled  longitudinal pressure, but not for the shear and bulk viscous corrections separately.  We further present evidence that both late- and early-time attractors exist for all moments of the one-particle distribution function that contain greater than one power of the longitudinal momentum squared.
}

\keywords{non-conformal relativistic kinetic theory, relativistic dissipative hydrodynamics, hydrodynamic attractors, non-equilibrium attractors, quark-gluon plasma}

\begin{document} 
\maketitle
\flushbottom

\section{Introduction}
\label{sect:intro}

One of the most important questions that has emerged in the last twenty years in the area of far-from-equilibrium relativistic dynamics is to what extent can such dynamics be described by relativistic viscous hydrodynamics.  In this context, the construction of a set of exact solutions to the relativistic Boltzmann equation in relaxation time approximation (RTA), despite its relative simplicity, has proven to be very useful in assessing the quantitative reliability of different dissipative hydrodynamical frameworks \cite{Florkowski:2013lza,Florkowski:2013lya,Florkowski:2014sfa,Florkowski:2017jnz,Denicol:2014xca,Denicol:2014tha}.  In addition, these exact solutions have helped to understand the emergence of a far-from-equilibrium attractor in relativistic transport theory that matches smoothly onto viscous hydrodynamics at late times but which extends to earlier times when conventional linearized viscous hydrodynamics treatments break down \cite{Heller:2015dha,Keegan:2015avk,Heller:2016rtz,Florkowski:2017olj,Romatschke:2017vte,Spalinski:2017mel,Romatschke:2017acs,Behtash:2017wqg,Florkowski:2017ovw,Strickland:2017kux,Almaalol:2018ynx,Denicol:2018pak,Behtash:2018moe,Strickland:2018ayk,Strickland:2019hff,Behtash:2019qtk,Behtash:2019txb,Brewer:2019oha,Blaizot:2020gql,Du:2020zqg,Du:2020dvp,Kamata:2020mka,Dore:2020jye,Blaizot:2021cdv,Soloviev:2021lhs,Alqahtani:2022xvo,Chattopadhyay:2021ive,Jaiswal:2022udf,Chattopadhyay:2022sxk,Kamata:2022jrc,Du:2022bel,Brewer:2022vkq}.  

Although the majority of these references focused on conformal systems, some of these works have considered whether or not attractors exist in non-conformal systems since in this case more than one dimensionful scale appears in the problem~\cite{Romatschke:2017acs,Florkowski:2017jnz,Chattopadhyay:2021ive,Jaiswal:2021uvv,Jaiswal:2022udf,Chattopadhyay:2022sxk}.  In this work we consider exact solutions of the RTA Boltzmann equation for a massive gas using the exact solution obtained originally in ref.~\cite{Florkowski:2014sfa}.  We extend this solution to allow computation of all moments, ${\cal M}^{nl}$, of the one-particle distribution using the moments introduced in ref.~\cite{Strickland:2018ayk}.  We also extend prior works by making use of the self-consistently determined temperature and mass dependent relaxation time, $\tau_{\rm eq}(T,m)$, and considering fixed specific shear viscosity.  In refs.~\cite{Romatschke:2017acs,Florkowski:2017jnz,Chattopadhyay:2021ive,Jaiswal:2021uvv,Jaiswal:2022udf,Chattopadhyay:2022sxk} either a constant relaxation time or conformal relaxation time proportional to the inverse temperature was used.  Finally, we systematically study both the forward and pull-back (early-time) attractors by varying both the initial anisotropy and initialization time and computing a large set of integral moments of the distribution function.

We will demonstrate that kinetic theory with an RTA collisional kernel possesses both forward and pull-back attractor for moments containing greater than one integral power of the longitudinal momentum squared ($l \geq 1$).  The existence of a forward attractor for such moments is established by holding the initialization time and energy density fixed while varying the initial momentum-space anisotropy using a spheroidal form for the initial one-particle distribution function.  Secondly, we establish the existence of an early-time (pull-back) attractor for such moments by holding the initial anisotropy and energy density fixed while varying the initialization time.  As we will demonstrate, this implies that there does not exist an early time attractor for the pressure-scaled shear and bulk viscous corrections independently, however, the difference of the two does possess an attractor, which is consistent with there being an attractor in the scaled longitudinal pressure.  Our findings are fully compatible with and extend those reported in ref.~\cite{Chattopadhyay:2021ive,Jaiswal:2021uvv}.

The structure of this paper is as follows.  In sec.~\ref{sect:setup}, we review basic thermodynamic and dynamic relations for non-conformal massive gases and extend the exact solution of the RTA Boltzmann equation obtained in ref.~\cite{Florkowski:2014sfa} to all moments of the distribution function.  In sec.~\ref{sect:dissipativehydro} we collect analytic formula for the viscosity corrected distribution functions and moments which are accurate to first order in hydrodynamic gradients using both the 14-moment and Chapman-Enskog approximations.  In sec.~\ref{sect:results}, we present our numerical results obtained from the exact solution of the RTA Boltzmann equation.  In Sec.~\ref{sect:conclusions}, we present our conclusions and an outlook for the future. 

\section{Setup}
\label{sect:setup}

In this work we will make use of a previously obtained exact solution to the 0+1d RTA Boltzmann equation for a massive gas with Boltzmann statistics.  The non-conformal exact solution was first presented in ref.~\cite{Florkowski:2014sfa} and extended earlier conformal exact solutions in a Bjorken expansion scenario~\cite{Florkowski:2013lza,Florkowski:2013lya}.  We will extend the original work of ref.~\cite{Florkowski:2014sfa} to include a relaxation time that self-consistently depends on both the temperature and mass of the particle, whereas the original work considered a constant relaxation time or the limit of low temperatures.  Our solution also goes beyond the considerations of the recent work of refs.~\cite{Romatschke:2017acs,Chattopadhyay:2021ive,Chattopadhyay:2022sxk} where a conformal relaxation time was used.

\subsection{RTA Boltzmann equation}
\label{sect:rta}

All results presented herein follow from the RTA Boltzmann equation in relaxation time approximation
\begin{equation}
 p^\mu \partial_\mu  f(x,p) =  C[ f(x,p)] \, , 
\label{kineq}
\end{equation}
where $f$ is the one-particle distribution function, $p^\mu$ is the particle four-momentum, and $C$ is the collision kernel 
\begin{eqnarray}
C[f] = \frac{p \cdot u}{\tau_{\rm eq}} \left( f_{\rm eq}-f \right) ,
\label{col-term}
\end{eqnarray}
with $u^\mu$ being the four-velocity of the local rest frame and $a \cdot b \equiv a^\mu b_\mu$.  The quantity $\tau_{\rm eq}$ appearing above is the relaxation time, which will be precisely specified below.  For the equilibrium distribution, we will follow ref.~\cite{Florkowski:2014sfa} and assume a Boltzmann distribution\footnote{It is possible to investigate the emergence of attractors with underlying Fermi-Dirac or Bose statistics using the exact solution presented in ref.~\cite{Florkowski:2014sda}.  We postpone this to future work.}
\begin{eqnarray}
f_{\rm eq} = \exp\left(- \frac{p \cdot u}{T} \right) .
\label{Boltzmann}
\end{eqnarray}
Herein, we will assume Bjorken flow, in which case in Milne coordinates one has $u^\tau=1$ and $u^{x,y,\varsigma}=0$, where $\tau $ is the longitudinal proper-time, $\tau = \sqrt{t^2-z^2}$, and $\varsigma$ is the spatial rapidity, $\varsigma = \tanh^{-1}(z/t)$.

\subsection{Thermodynamic variables}
\label{sect:thermo}

For a single-component massive gas obeying Boltzmann statistics, the equilibrium thermodynamic quantities are
\ba
{\mathpzc n} &=& \frac{T^3}{2 \pi^2} \, \hat{m}^2 K_2\left( \hat{m}\right) , \label{eq:neq} \nonumber \\
s &=& \frac{T^3}{2 \pi^2} \, \hat{m}^2 \Big[4K_2\left( \hat{m}\right)+\hat{m}K_1\left( \hat{m}\right)\Big] ,
\label{eq:Seq} \nonumber \\
\varepsilon &=& \frac{T^4}{2 \pi^2} \, \hat{m}^2
 \Big[ 3 K_{2}\left( \hat{m} \right) + \hat{m} K_{1} \left( \hat{m} \right) \Big]  \, , 
\label{eq:Eeq} \nonumber \\
 P &=& {\mathpzc n} T = \frac{T^4}{2 \pi^2} \, \hat{m}^2 K_2\left( \hat{m}\right)  \, ,
\label{eq:Peq}
\ea
with $\hat{m} \equiv m/T$ and $K_n$ being modified Bessel functions of the second kind.  Above $\mathpzc n$ is the number density, $s$ is the entropy density, $\varepsilon$ is the energy density, and $P$ is the pressure.  These satisfy $\varepsilon + P = Ts$ and, from the above relations, one can determine the speed of sound squared
\be
c_s^2 = \frac{dP}{d\varepsilon} = \frac{\varepsilon+P}{3\varepsilon+(3+\hat{m}^2)P} \, .
\ee

\subsection{Relaxation time for a massive gas}
\label{sect:relaxationtime}

\begin{figure}[ht]
\centerline{\includegraphics[width=0.5\linewidth]{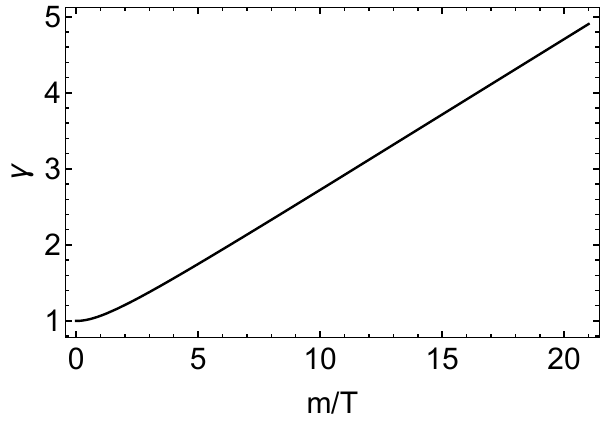}}

\caption{The non-conformal relaxation time modification factor $\gamma$ \eqref{eq:gammadef} as a function of $m/T$. }
\label{fig:teqfac}
\end{figure}

For a massive system, the shear viscosity $\eta$ can be expressed as \cite{anderson1974relativistic,Czyz:1986mr,Alqahtani:2015qja}
\be
\eta = \frac{\tau_{\rm eq} P}{15}\kappa(\hat{m})\, ,
\ee
with
\be 
\kappa(x)\equiv x^3 \bigg[\frac{3}{x^2}\frac{K_3(x)}{K_2(x)}-\frac{1}{x}+\frac{K_1(x)}{K_2(x)}-\frac{\pi}{2}\frac{1-xK_0(x)L_{-1}(x)-xK_1(x)L_0(x)}{K_2(x)}\bigg] \, ,
\ee
and $L_n(x)$ being modified Struve functions. 
For fixed specific shear viscosity, $\bar\eta \equiv \eta/s$, using $\varepsilon+P=Ts$ one obtains
\be 
\tau_{\rm eq}(T,m)=\frac{5 \bar{\eta}}{T} \gamma(\hat{m}) \, ,
\label{eq:teq2}
\ee
with 
\be
\gamma(\hat{m}) \equiv \frac{3}{\kappa(\hat{m})} \bigg(1+\frac{\varepsilon}{P}\bigg) \, .
\label{eq:gammadef}
\ee
Note that, in the massless limit, $m\rightarrow 0$, one has $\kappa(\hat{m})\rightarrow12$, $\varepsilon \rightarrow 3P$, and $\gamma \rightarrow 1$, giving the usual conformal RTA relaxation time 
\be
\tau_{\rm eq}(T,0)= \frac{5 \bar{\eta}}{T} \, .
\label{eq:teq0}
\ee
For small $\hat{m}$, one has
\be 
\gamma(\hat{m}) = 1+\frac{\hat{m}^2}{12}-\frac{13 \hat{m}^4}{288}+{\cal O}\!\left(\hat{m}^5\right),
\ee
and in the large $\hat{m}$ limit, one has
\be
\gamma(\hat{m}) = \frac{\hat{m}}{5} + \frac{7}{10} + {\cal O}\!\left(\frac{1}{\hat{m}}\right).
\ee
In fig.~\ref{fig:teqfac} we plot $\gamma(\hat{m})$.  As can be seen from this figure, $\gamma(\hat{m})$ goes to unity in the massless limit and grows linearly at large $m/T$, which corresponds either to fixed temperature and large mass or fixed mass and small temperature.  The fact that $\gamma(\hat{m}) \geq 1$ implies that a massive gas always relaxes more slowly to equilibrium than a massless one in physical units, however, it is unclear a priori how things will change as a function of the rescaled time $\wbar \equiv \tau/\tau_{\rm eq}$.  We note that the strong enhancement of the relaxation time at low temperatures modifies the asymptotic approach to equilibrium.

\subsection{Exact solution for the distribution function and its solution}
\label{sect:exactsolf}

In this section we review the derivation of the exact solution presented in ref.~\cite{Florkowski:2014sfa} and derive the integral equation obeyed by all moments of the distribution function.  We also generalize the results contained in that reference to the full set of integral moments.

In the case of one-dimensional boost-invariant expansion (0+1d), all scalar quantities depend only on the longitudinal proper time $\tau$.  To describe boost-invariant 0+1d dynamics, one can introduce a spacelike vector $z^\mu$, which is orthogonal to the fluid four-velocity $u^\mu$ in all frames and corresponds to the z-direction in the local rest frame of the matter \cite{Ryblewski:2010ch,Martinez:2012tu}.

The requirement of boost invariance implies that $f(x,p)$ may depend only on three variables, $\tau$, $w$, and $\vec{p}_T$ \cite{Bialas:1984wv,Bialas:1987en}, with the boost-invariant variable $w$ defined by\footnote{In eq.~\eqref{eq:w}, $z$ is the spatial coordinate, which is not to be confused with the basis vector $z^\mu$.}
\be
w =  t p_L - z E \, .
\label{eq:w}
\ee
Using $w$ and $\vec{p}_T$ one can define
\be
v = Et-p_L z = 
\sqrt{w^2+\left( m^2+\vec{p}_T^{\,\,2}\right) \tau^2} \, .
\label{eq:v}
\ee
Using these variables, one can write the energy and the longitudinal momentum as
\ba
E &=& \frac{vt+wz}{\tau^2} \, , \\
p_L &=& \frac{wt+vz}{\tau^2} \, ,
\label{eq:p0p3}
\ea
and the Lorentz-invariant momentum integration measure becomes
\begin{equation}
dP =   \frac{d^4p}{(2\pi)^4} \, 2\pi \delta \left( p^2-m^2\right) 2 \theta (p^0)
=\frac{dp_L}{(2\pi)^3p^0}d^2p_T =\frac{dw \, d^2p_T }{(2\pi)^3v}\, .  
\label{dP}
\end{equation}

When written in terms of these variables, the 0+1d RTA Boltzmann equation takes a particularly simple form~\cite{Florkowski:2013lza,Florkowski:2013lya,Florkowski:2014sfa}
\begin{eqnarray}
\frac{\partial f}{\partial \tau}  &=& 
\frac{f_{\rm eq}-f}{\tau_{\rm eq}} \, ,
\label{eq:simpleformrta}
\end{eqnarray} 
with $\tau_{\rm eq}$ specified in eq.~\eqref{eq:teq2} and the equilibrium distribution function given by
\begin{eqnarray}
f_{\rm eq}(\tau,w,p_T) =
\exp\!\left[
- \frac{\sqrt{w^2+ (p_T^2+m^2)\tau^2}}{T \tau}  \right] .
\label{eqdistform}
\end{eqnarray}
The exact solution to eq.~\eqref{eq:simpleformrta} is~\cite{Florkowski:2013lza,Florkowski:2013lya,
Baym:1984np,Baym:1985tna,Heiselberg:1995sh,Wong:1996va,Florkowski:2014sfa}
\begin{equation}
f(\tau,w,p_T) = D(\tau,\tau_0) f_0(w,p_T)  + \int_{\tau_0}^\tau \frac{d\tau^\prime}{\tau_{\rm eq}(\tau^\prime)} \, D(\tau,\tau^\prime) \, 
f_{\rm eq}(\tau^\prime,w,p_T) \, ,  \label{eq:solf}
\end{equation}
where $f_0(w,p_T)$ is the initial distribution function specified at $\tau = \tau_0$ and the damping function $D$ is defined as
\begin{equation}
D(\tau_2,\tau_1) = \exp\left[-\int_{\tau_1}^{\tau_2}
\frac{d\tau^{\prime\prime}}{\tau_{\rm eq}(\tau^{\prime\prime})} \right].
\end{equation}

In the main body of this work, we will assume that the initial distribution function $f_0$ can be expressed in spheroidally-deformed form~\cite{Romatschke:2003ms,Romatschke:2004jh}
\begin{eqnarray}
f_0(w,p_T) &=& 
\exp\left[
-\frac{\sqrt{(p\cdot u)^2 + \xi_0 (p\cdot z)^2}}{\Lambda_0} \, \right] \nonumber \\
&=& 
\exp\left[
-\frac{\sqrt{(1+\xi_0) w^2 + (m^2+p_T^2) \tau_0^2}}{\Lambda_0 \tau_0}\, \right] ,
\label{G0}
\end{eqnarray}
where $\xi_0$ is the initial anisotropy parameter and $\Lambda_0$ is the initial transverse momentum scale.  We consider a slight generalization of this initial condition in app.~\ref{app:a}.
For $-1 < \xi_0 <0$, this corresponds to an initially prolate distribution in the local rest frame and, conversely, for $\xi_0 > 0$ this corresponds to an initially oblate distribution function. For $\xi_0 = 0$, one obtains an isotropic Boltzmann distribution function as the initial condition.

\subsection{The integral equation obeyed by all moments}
\label{sect:moments}

We will work with the following moments of the one-particle distribution function~\cite{Strickland:2018ayk,Strickland:2019hff}
\be
{\cal M}^{nl}[f] \equiv \int dP \,(p \cdot u)^n \, (p \cdot z)^{2l} \, f(\tau,w,p_T) \, .
\label{eq:genmom1}
\ee
In principle, powers of $p_T^2$ could also appear in a general moment, however, such moments can be expressed as a linear combination of the two-index moment appearing above using $p^2=m^2$ to write $p_T^2 = (p \cdot u)^2 - (p\cdot z)^2 - m^2$.

Some specific cases of ${\cal M}^{nl}$ map to familiar quantities, e.g., $n=1$ and $l=0$ maps to the number density ${\mathpzc n} = {\cal M}^{10}$, $n=2$ and $l=0$ maps to the energy density, and $n=0$ and $l=1$ maps to the longitudinal pressure, $P_L$.  The transverse pressure, $P_T$, can be obtained by using $p_T^2 = (p \cdot u)^2 - (p\cdot z)^2 - m^2$ to obtain $P_T = {\cal M}^{20} - {\cal M}^{01} - m^2 {\cal M}^{00}$.

For a Boltzmann equilibrium distribution function, these moments reduce to 
\be
{\cal M}^{nl}_{\rm eq}(T,m) \equiv {\cal M}^{nl}[f_{\rm eq}] = \frac{2 T^{n+2l+2}}{(2\pi)^2(2l+1)} \int_0^\infty d\hat{p} \, \hat{p}^{n+2l+1} \left( 1 + \frac{\hat{m}^2}{\hat{p}^2} \right)^{(n-1)/2}  e^{-\sqrt{\hat{p}^2+\hat{m}^2}} \, .
\ee
We note that by changing variables to $x \equiv \sqrt{\hat{p}^2+\hat{m}^2}$ it is possible to perform this integral analytically in terms of generalized hypergeometric functions, however, the resulting expression is not straightforward to evaluate for integer-valued $n$ and $l$.  For this reason, it is typically easier to simply evaluate it numerically.

In what follows, we will present results for these general moments scaled by their equilibrium values, i.e., 
\be
{\Mbar}^{nl} \equiv \frac{{\cal M}^{nl}}{{\cal M}^{nl}_{\rm eq}} \, .
\ee
In the late-time limit ($\tau \rightarrow \infty$), if the system approaches equilibrium, then ${\Mbar}^{nl} \rightarrow 1$.

In the general case, using the boost-invariant variables introduced earlier, one finds that the general moments can be expressed as
\ba
{\cal M}^{nl}[f] &=& \int  \frac{dw \, d^2p_T }{(2\pi)^3v} \left( \frac{v}{\tau} \right)^n \left( \frac{w}{\tau} \right)^{2l} \, f(\tau,w,p_T) \, , \nonumber \\
&=& \frac{1}{(2\pi)^3 \, \tau^{n+2l}} \int  dw \, d^2p_T  \, v^{n-1} w^{2l} \, f(\tau,w,p_T)
\ea
Taking a general moment of eq.~(\ref{eq:solf}) one obtains
\be
{\cal M}^{nl}(\tau) = D(\tau,\tau_0) {\cal M}^{nl}_0(\tau)  + \int_{\tau_0}^\tau \frac{d\tau^\prime}{\tau_{\rm eq}(\tau^\prime)} \, D(\tau,\tau^\prime) \, 
{\cal M}^{nl}_{\rm eq}(\tau') .  \nonumber
\ee

Evaluating the integrals necessary results in
\ba
{\cal M}^{nl} &=&  \frac{D(\tau,\tau_0) \Lambda_0^{n+2l+2}}{(2\pi)^2}\tilde{H}^{nl}\left(\frac{\tau_0}{\tau \sqrt{1+\xi_0}}, \frac{m}{\Lambda_0} \right) \nonumber \\
&& \hspace{2cm} + \frac{1}{(2\pi)^2} \int_{\tau_0}^\tau \frac{d\tau^\prime}{\tau_{\rm eq}(\tau^\prime)} \, D(\tau,\tau^\prime) \, 
T^{n+2l+2}(\tau^\prime) \, \tilde{H}^{nl}\left(\frac{\tau'}{\tau}, \frac{m}{T(\tau^\prime)} \right)  \, , \label{eq:mnleq}
\ea
where
\be
\tilde{H}^{nl}(y,z) =
\int_{0}^{\infty} du \, u^{n+2l+1} e^{-\sqrt{u^2 + z^2}} \,  H^{nl}\!\left(y,\frac{z}{u}\right) ,
\ee
with
\be
H^{nl}(y,x) = \frac{2\,y^{2l+1}
   (1+x^2)^{\frac{n-1}{2}}}{2l+1}
   \,_2F_1\!\left(l+\frac{1}{2},\frac{1-n}{2};l
   +\frac{3}{2};\frac{1-y^2}{1+x^2}\right),
\ee
where $_2F_1$ is a hypergeometric function.
Finally, specializing to the case $n=2$ and $l=0$ and requiring conservation of energy $\varepsilon(\tau) =\varepsilon_{\rm eq}(T)$, also known as Landau matching, we obtain the following integral equation
\ba
&& \hspace{-1cm} 2 T^4(\tau) \, \hat{m}^2
 \left[ 3 K_2\!\left( \frac{m}{T(\tau)} \right) + \hat{m} K_1\!\left( \frac{m}{T(\tau)} \right) \right]
 \nonumber \\ 
 && = D(\tau,\tau_0) \Lambda_0^4 \tilde{H}^{20}\!\left(\frac{\tau_0}{\tau\sqrt{1{+}\xi_0}},\frac{m}{\Lambda_0}\right) +  \int_{\tau_0}^\tau \frac{d\tau^\prime}{\tau_{\rm eq}(\tau^\prime)} \, D(\tau,\tau^\prime) \, 
T^4(\tau^\prime) \tilde{H}^{20}\!\left(\frac{\tau'}{\tau}, \frac{m}{T(\tau^\prime)} \right) .
\label{eq:20final}
\ea
This is the integral equation obtained originally in ref.~\cite{Florkowski:2014sfa} with the understanding that \mbox{$\tilde{H}^{20} = \tilde{\cal H}_2$} defined therein.  This equation can be solved iteratively for $T(\tau)$ and, once converged to the desired numerical accuracy, the solution can be used in eq.~\eqref{eq:mnleq} to compute all moments.

\subsection{Viscous corrections expressed in terms of moments}
\label{sect:viscousmoments}

For comparisons to come, here we collect expressions for the viscous corrections written in terms of the moments computed herein.  We start by noting that the equilibrium pressure can be expressed as
\be
P = - \frac{1}{3}\Delta_{\mu\nu} \int dP \, p^\mu p^\nu \, f_{\rm eq} = \frac{1}{3} \left[ {\cal M}^{20}_{\rm eq} - m^2 {\cal M}^{00}_{\rm eq} \right] \, .
\ee
where $\Delta_{\mu\nu}$ is the projection operator onto the orthogonal to $u^\mu$.
Next, we note that the bulk viscous correction can be expressed as
\ba
\Pi &=& - \frac{1}{3}\Delta_{\mu\nu} \int dP \, p^\mu p^\nu \, (f-f_{\rm eq}) \nonumber \\
&=&\frac{1}{3} \left[ {\cal M}^{20} - m^2 {\cal M}^{00} \right] - \frac{1}{3} \left[ {\cal M}^{20}_{\rm eq} - m^2 {\cal M}^{00}_{\rm eq} \right] \nonumber \\
&=& - \frac{1}{3} m^2 \left[ {\cal M}^{00} - {\cal M}^{00}_{\rm eq} \right] ,
\ea
where, in going from the second the third lines we have used Landau matching, which implies that ${\cal M}^{20} = {\cal M}^{20}_{\rm eq}$.
Scaling by the equilibrium pressure, one obtains
\be
\tilde\Pi \equiv \frac{\Pi}{P} = -\frac{m^2 \left({\cal M}^{00} - {\cal M}^{00}_{\rm eq}\right)}{{\cal M}^{20}_{\rm eq} - m^2 {\cal M}^{00}_{\rm eq}} \, .
\label{eq:tildebulk}
\ee
From this we see that $\tilde\Pi$ is proportional to the difference of the $n=0$ and $l=0$ moment from its equilibrium value.  We note that, if ${\cal M}^{00}$ does not possess an attractor, this would imply that $\tilde\Pi$ does not possess an attractor.
To compute the shear correction, it is most straightforward to start from
\be
{\cal M}^{01} = P_L = P - \pi + \Pi \, ,
\ee
which results in
\be
\tilde\pi \equiv \frac{\pi}{P} = 1 - \Mbar^{01} + \tilde\Pi \, .
\label{eq:tildeshear}
\ee
We note that this implies that, if there exists an attractor for $\Mbar^{01}$, but not for $\tilde\Pi$, then $\tilde\pi$ will also not possess an attractor.  This point was originally emphasized in refs.~\cite{Chattopadhyay:2021ive,Jaiswal:2021uvv}.

\section{Evaluation of the moments to first order in hydrodynamic gradients}
\label{sect:dissipativehydro}

In this section we will present expressions for the shear and bulk viscosity corrected distribution functions and resulting integral moments obtained using both the 14-moment~\cite{grad_1949,Denicol:2010xn,Denicol:2011fa} and Chapman-Enskog~\cite{Chapman1991-qu,Jaiswal:2014isa} approximations.  In both cases, one can decompose the linearly-corrected one-particle distribution function as
\be
f = f_{\rm eq} + \delta f_{\rm shear} + \delta f_{\rm bulk} \, .
\label{eq:vfreeze}
\ee
In the next two subsections we specify the 14-moment and Chapman-Enskog forms for $\delta f_{\rm shear}$ and $\delta f_{\rm bulk}$ and evaluate the moments of each in order to obtain the corresponding approximations at first order in gradients.

\subsection{14-moment approximation}

In the 14-moment approximation, the viscous corrections to the distribution function for a single component massive gas obeying classical statistics can be written as~\cite{Teaney:2003kp,Bozek:2009dw,Rose:2014fba,Alqahtani:2016rth}
\ba
\delta f_{\rm shear} &=& f_{\rm eq} \frac{p_\mu p_\nu \pi^{\mu\nu}}{2(\varepsilon+P)T^2} \, , 
\label{eq:deltashear} \\
\delta f_{\rm bulk} &=& - f_{\rm eq} \frac{\beta}{\beta_\Pi} \bigg[ \frac{m^2}{3\,  p \cdot u } - \Big(\frac{1}{3}-c_s^2\Big) p \cdot u \bigg] \Pi \, , 
\label{eq:deltafbulk}
\ea
with $\beta = 1/T$ and
\be
\beta_\Pi = \frac{5}{3}\beta\, I_{42}^{(1)} - (\varepsilon+P)c_s^2 \, .
\ee
The thermodynamic integral $I_{42}^{(1)}$ can be expressed as~\cite{Jaiswal:2014isa}
\be
I_{42}^{(1)} = \frac{T^5\hat{m}^5}{30\pi^2}\left[\frac{1}{16}\Big(K_5(\hat{m})-7K_3(\hat{m})+22K_1(\hat{m})\Big)-K_{i,1}(\hat{m})\right] ,
\label{eq:i142}
\ee
with 
\begin{equation}\label{kin}
K_{i,1}(\hat{m}) = \int_0^\infty\! \frac{d\theta}{\cosh\theta} \,\exp(-\hat{m}\cosh\theta) \, .
\end{equation}
For 0+1d boost-invariant Bjorken expansion as considered herein, one can write the shear tensor in terms of one independent component $\pi \equiv - \pi^{zz}$, with the other two diagonal components determined by symmetry and the tracelessness of $\pi^{\mu\nu}$, giving $\pi^{xx} = \pi^{yy} = \pi/2$.  Note that in this case, all dynamical variables only depend on the longitudinal proper time $\tau$.

\subsubsection*{Navier-Stokes shear-viscous correction}

For the case of 0+1d boost-invariant Bjorken expansion one has
\be
\delta f_{\rm shear,14-moment} = \frac{ \bar\pi}{4\left(1+ \frac{P}{\varepsilon}\right) T^2} \left[ (p\cdot u)^2 - 3 (p\cdot z)^2 - m^2 \right] f_{\rm eq}  \, ,
\ee
where $\bar\pi = \pi/\varepsilon$.  Computing the moments of $\delta f_{\rm shear}^\text{14-moment}$ one obtains
\be
{\cal M}^{nl}_{\rm shear,14-moment} = \frac{ \bar\pi}{4\left(1+ \frac{P}{\varepsilon}\right)  T^2} \left[  {\cal M}_{\rm eq}^{n+2,l} - 3 {\cal M}_{\rm eq}^{n,l+1} - m^2 {\cal M}_{\rm eq}^{n,l} \right] ,
\ee
At first order in the gradient expansion, which corresponds to the Navier-Stokes (NS) limit, one has
\be
\pi^\text{NS}(\tau) = \frac{4\eta}{3\tau} \, ,
\label{eq:shearns}
\ee
and $\bar\pi^{\rm NS} = \frac{4 \bar\eta}{3 \tau T}\left(1+ \frac{P}{\varepsilon}\right) $, which when written in terms of $\wbar = \tau/\tau_{\rm eq}$, becomes
\be
\bar\pi^{\rm NS}  = \frac{4}{15 \gamma(\hat{m}) \wbar} \left( 1 + \frac{P}{\varepsilon} \right) \, ,
\ee
giving
\be
{\cal M}^{nl,\rm NS}_{\rm shear,14-moment} = \frac{1}{15 \, \wbar \, T^2 \, \gamma(\hat{m})} \left[  {\cal M}_{\rm eq}^{n+2,l} - 3 {\cal M}_{\rm eq}^{n,l+1} - m^2 {\cal M}_{\rm eq}^{n,l} \right] .
\ee

\subsubsection*{Navier-Stokes bulk-viscous correction}

Taking the moments of eq.~\eqref{eq:deltafbulk}, one obtains
\be
{\cal M}^{nl}_{\rm bulk,14-moment} = - \frac{\beta}{3 \beta_\Pi} \left[ m^2 
{\cal M}_{\rm eq}^{n-1,l} - \left(1-3c_s^2\right) {\cal M}_{\rm eq}^{n+1,l} \right] \Pi \, .
\ee
To proceed, one can use the fact that $\Pi = - \, \tau_{\rm eq} \, \beta_\Pi \, \partial_\mu u^\mu$ \cite{Jaiswal:2014isa}.  At first order, for boost-invariant Bjorken flow, since $\partial_\mu u^\mu = 1/\tau$, this reduces to
\be
\Pi^{\rm NS} = - \beta_\Pi/\wbar \, ,
\label{eq:bulkns}
\ee
giving
\be
{\cal M}^{nl,\rm NS}_{\rm bulk,14-moment} = \frac{1}{3 \, \wbar \, T } \left[ m^2 
{\cal M}_{\rm eq}^{n-1,l} - \left(1-3c_s^2\right) {\cal M}_{\rm eq}^{n+1,l} \right] .
\label{eq:14bulkns}
\ee

\subsubsection*{Total Navier-Stokes viscous correction}

Finally, by adding the shear and bulk corrections to the equilibrium result and scaling by the equilibrium moments, we obtain the following expression for the scaled moments in the Navier-Stokes limit within the 14-moment approximation
\ba
\Mbar^{nl,\rm NS}_{\rm 14-moment} &=& 1 + \frac{1}{15 \, \wbar \, T^2 \, \gamma(\hat{m})}\frac{\left[  {\cal M}_{\rm eq}^{n+2,l} - 3 {\cal M}_{\rm eq}^{n,l+1} - m^2 {\cal M}_{\rm eq}^{n,l} \right]}{{\cal M}_{\rm eq}^{n,l}} \nonumber \\
&& \hspace{4cm} + \frac{1}{3 \, \wbar \, T } \frac{\left[ m^2 
{\cal M}_{\rm eq}^{n-1,l} - \left(1-3c_s^2\right) {\cal M}_{\rm eq}^{n+1,l} \right]}{{\cal M}_{\rm eq}^{n,l}} \, . \;\;\;
\label{eq:14momfinal}
\ea

\subsection{Chapman-Enskog approximation}

\subsubsection*{Navier-Stokes shear-viscous correction}

In the Chapman-Enskog approximation one has the following shear viscous correction~\cite{Jaiswal:2014isa}
\be
\delta f_{\rm shear, CE} = \frac{\beta f_{\rm eq}}{2(u\cdot p)\beta_\pi}\;p^\mu p^\nu\pi_{\mu\nu} \, ,
\ee
where $\beta_\pi =  \beta\, I_{42}^{(1)}$.  Following a similar procedure as was used for the 14-moment approximation, ones find that, in the Navier-Stokes limit, the moments of the shear viscous correction become
\be
{\cal M}^{nl,\rm NS}_{\rm shear,CE} = \frac{\varepsilon + P}{15 \, \wbar \, \gamma(\hat{m}) \, I_{42}^{(1)} } \left[  {\cal M}_{\rm eq}^{n+1,l} - 3 {\cal M}_{\rm eq}^{n-1,l+1} - m^2 {\cal M}_{\rm eq}^{n-1,l} \right] ,
\ee
with $I_{42}^{(1)}$ given in eq.~\eqref{eq:i142} and $\gamma(\hat{m})$ given in eq.~\eqref{eq:gammadef}.

\subsubsection*{Navier-Stokes bulk-viscous correction}

The bulk viscous correction in the Chapman-Enskog approximation can be written as~\cite{Jaiswal:2014isa}
\be
\delta f_{\rm bulk, CE} = - \frac{\beta f_{\rm eq}}{3(u\cdot p)\beta_\Pi}\left[m^2-\left(1-3c_s^2\right)(u\cdot p)^2\right]\Pi \, , 
\ee
which is precisely the same form as the 14-moment approximation and hence the moments reduce to eq.~\eqref{eq:14bulkns}.

\subsubsection*{Total Navier-Stokes viscous correction}

Adding the shear and bulk corrections to the equilibrium result and scaling by the equilibrium moments, we obtain the following expression for the scaled moments in the Navier-Stokes limit within the Chapman-Enskog approximation
\ba
\Mbar^{nl,\rm NS}_{\rm CE} &=& 1 + \frac{\varepsilon + P}{15 \, \wbar \, \gamma(\hat{m}) \, I_{42}^{(1)} } \frac{\left[  {\cal M}_{\rm eq}^{n+1,l} - 3 {\cal M}_{\rm eq}^{n-1,l+1} - m^2 {\cal M}_{\rm eq}^{n-1,l} \right]}{{\cal M}_{\rm eq}^{n,l}} \nonumber \\
&& \hspace{4cm} + \frac{1}{3 \, \wbar \, T } \frac{\left[ m^2 
{\cal M}_{\rm eq}^{n-1,l} - \left(1-3c_s^2\right) {\cal M}_{\rm eq}^{n+1,l} \right]}{{\cal M}_{\rm eq}^{n,l}} \, . \;\;\;
\label{eq:cefinal}
\ea

\section{Results}
\label{sect:results}

The integral equation \eqref{eq:20final} can be solved iteratively for $T(\tau)$ and, once converged to the desired accuracy, this solution can be used in eq.~\eqref{eq:mnleq} to compute all moments.  For the iterative solution, we discretized $T(\tau)$ on a logarithmic grid in time with 4096 points.  We consider two cases:  (a) holding the initial energy density, $\varepsilon_0$, and initialization time, $\tau_0$, fixed, while varying the initial momentum anisotropy, $\xi_0$; and (b) holding the initial energy density, $\varepsilon_0$, and the initial momentum anisotropy, $\xi_0$ fixed, while varying the initialization time, $\tau_0$.  These two scenarios allow us to assess whether or not forward and early-time (or pull-back) attractors exist, respectively.  In both cases, we hold the specific shear viscosity $\bar\eta = \eta/s$ fixed during the entire evolution.\footnote{We consider a slight generalization of this initial condition in app.~\ref{app:a}.}

In both cases, the initial energy density used corresponds to a massive gas at a temperature of $T_0 = 1$ GeV and the final evolution time was held fixed at $\tau_f = 100$ fm/c. Additionally, in both cases we iterated the integral equation for 200 iterations, which allowed for convergence of the result to 8 digits at all proper times.  For case (a), we used $\tau_0 = 0.1$ fm/c and, for case (b), we used $\xi_0 = 0$.  We consider three constant masses of $m=0.2$ GeV, $m=1$ GeV, and $m=5$ GeV.  We note that we have explicitly checked that the small mass limit our results converge to the conformal limit presented in ref.~\cite{Strickland:2018ayk}.  The {\tt RTA-MASSIVE-CUDA} code used to generate all results can be obtained using ref.~\cite{MikeCodeDB}.

\begin{figure}[ht]
\centerline{\includegraphics[width=1\linewidth]{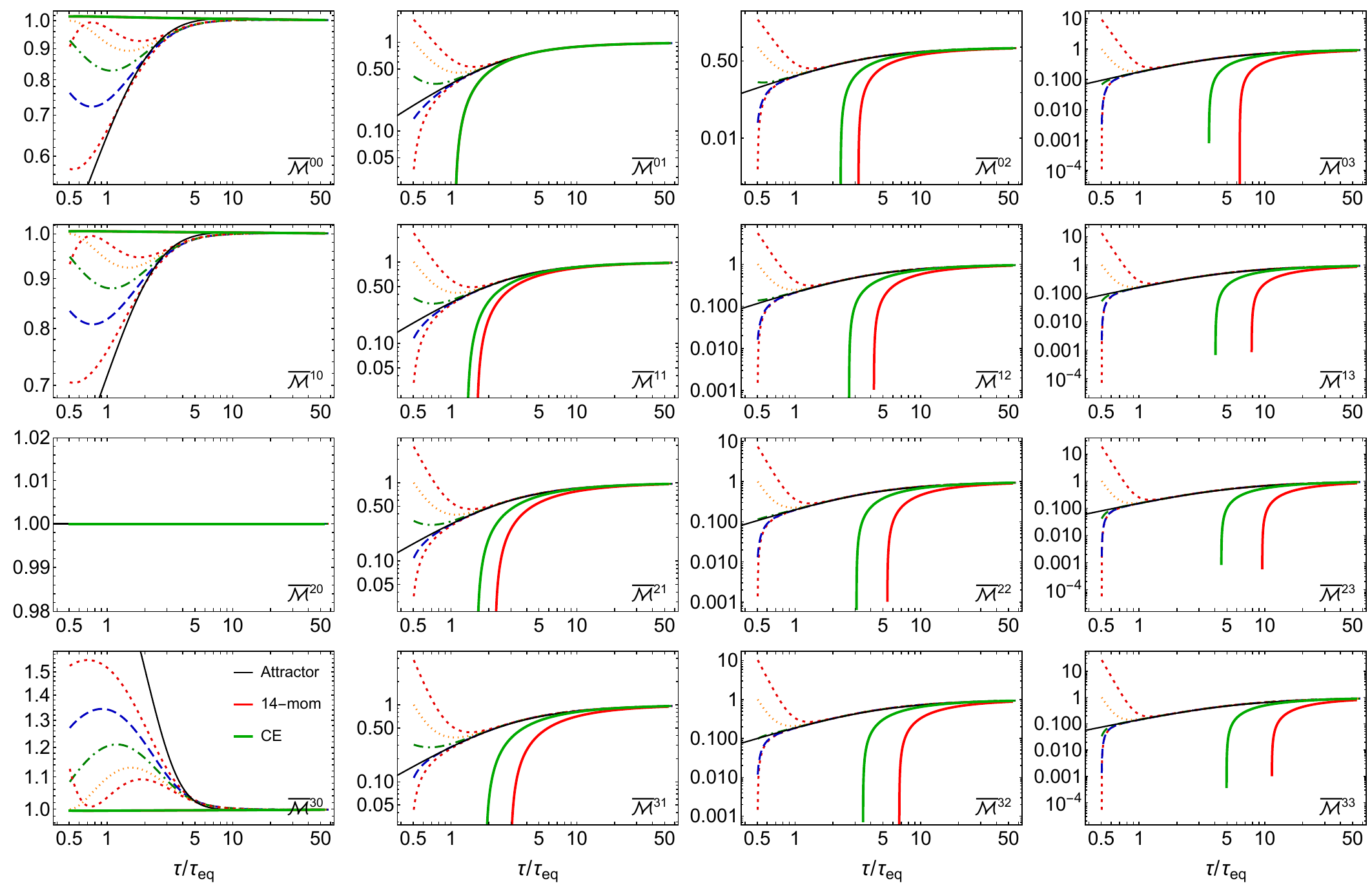}\hspace{2mm}}
\caption{Scaled moments $\Mbar^{nl}$ as a function of rescaled time for the case $m = 0.2$~GeV.  The non-solid lines are specific initial conditions initialized at $\tau_0 = 0.1$~fm/c with $T_0=1$~GeV and $\alpha_0 = 1/\sqrt{1+\xi_0} \in \{0.12,0.25,0.5,1,2\}$.  The solid black lines correspond to the attractor solution, the solid red lines are the first-order 14-moment predictions in eq.~\eqref{eq:14momfinal}, and the solid green lines are the first-order Chapman-Enskog predictions in eq.~\eqref{eq:cefinal}. }
\label{fig:a0scan-m0p2}
\end{figure}

\begin{figure}[ht]
\centerline{\includegraphics[width=1\linewidth]{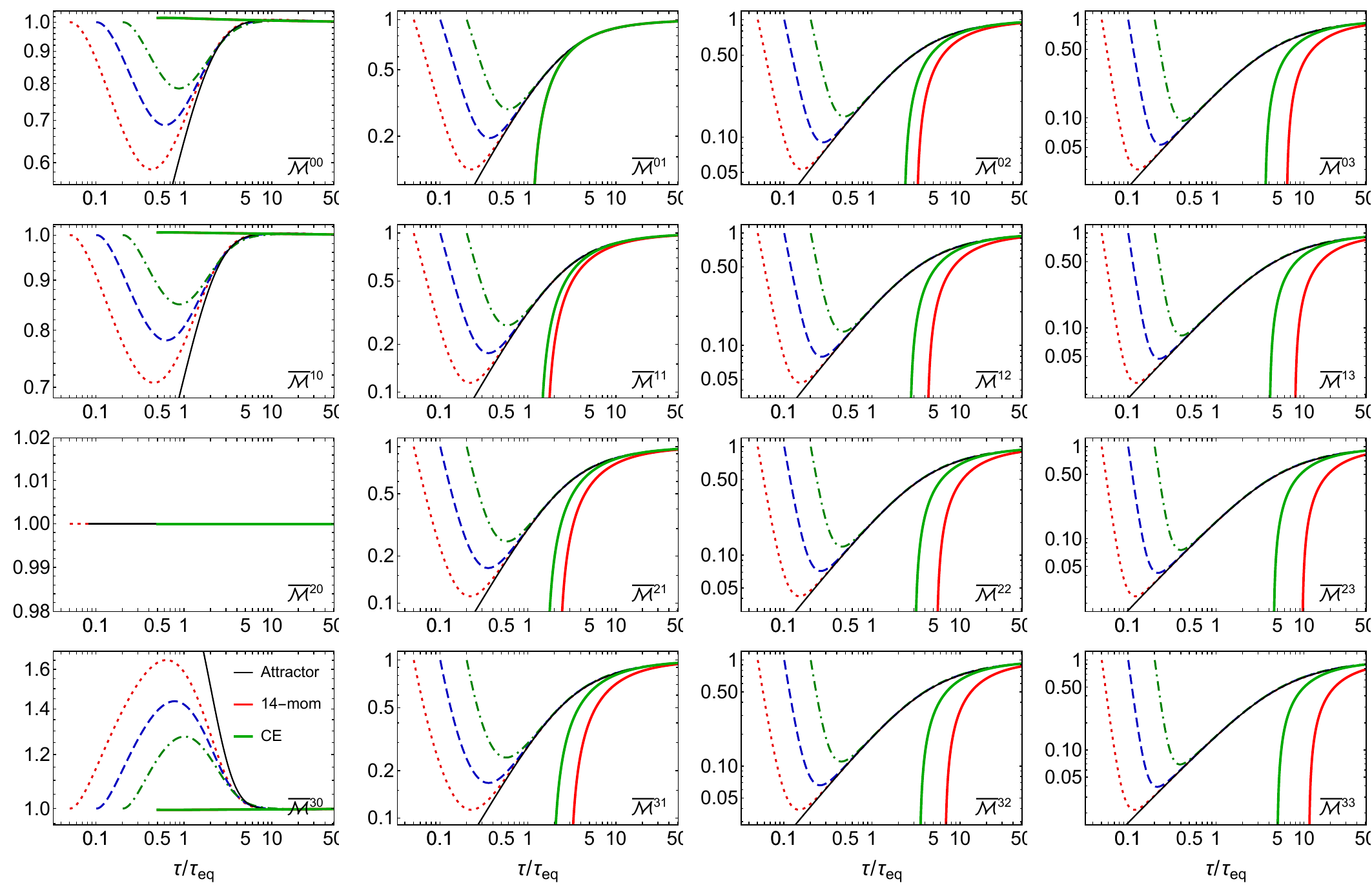}\hspace{2mm}}
\caption{Scaled moments $\Mbar^{nl}$ as a function of rescaled time for the case $m = 0.2$~GeV.  The non-solid lines are specific initial conditions initialized with $T_0 = 1$~GeV and $\xi_0=0$ at $\tau_0 \in \{0.01,0.02,0.04\}$~fm/c.  Line styles are the same as in fig.~\ref{fig:a0scan-m0p2}.  }
\label{fig:t0scan-m0p2}
\end{figure}

\begin{figure}[ht]
\centerline{\includegraphics[width=1\linewidth]{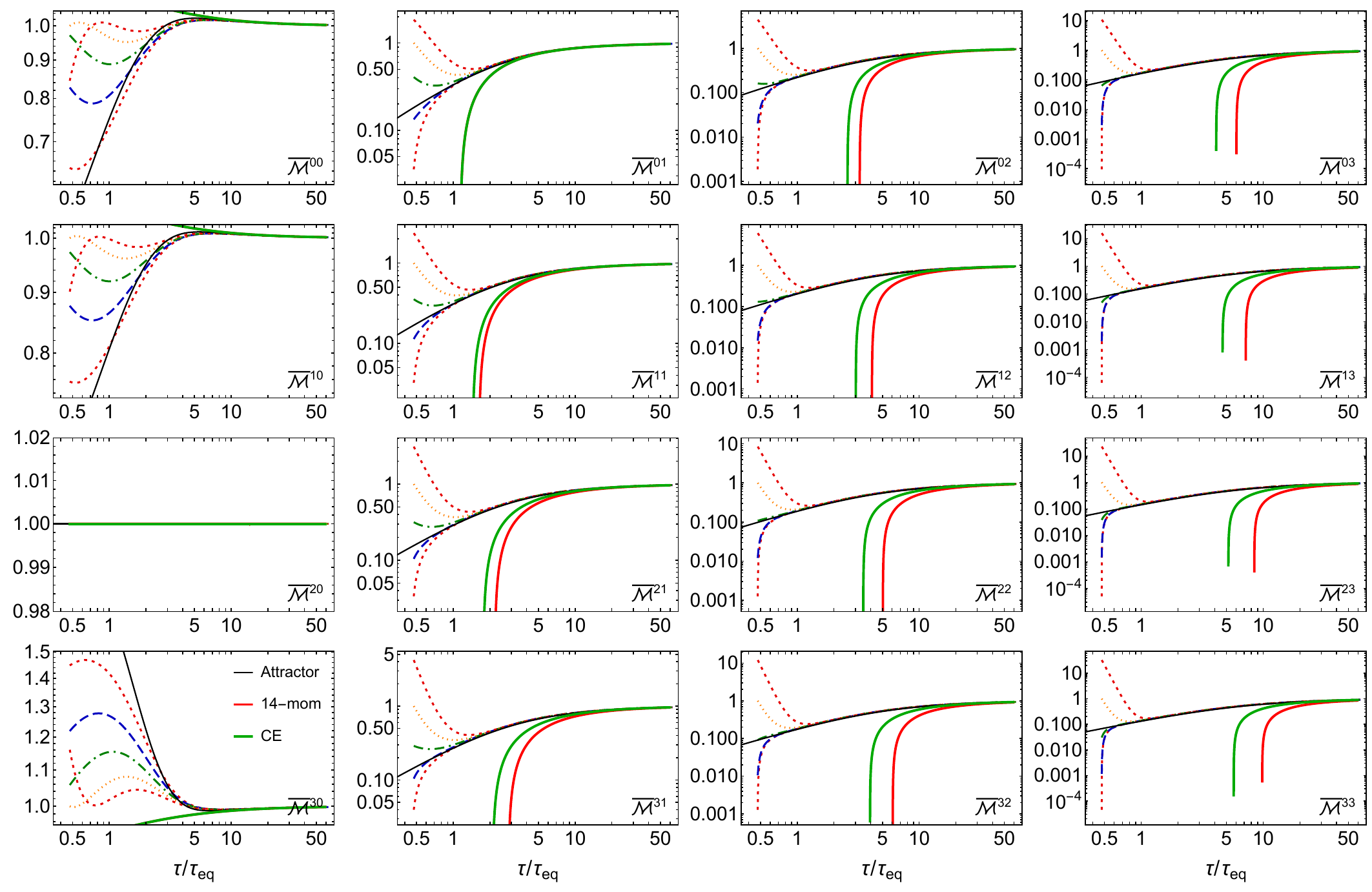}\hspace{2mm}}
\caption{Scaled moments $\Mbar^{nl}$ as a function of rescaled time for the case $m = 1$~GeV.  Initial conditions and line styles are the same as in fig.~\ref{fig:a0scan-m0p2}.}
\label{fig:a0scan-m1}
\end{figure}

\begin{figure}[ht]
\centerline{\includegraphics[width=1\linewidth]{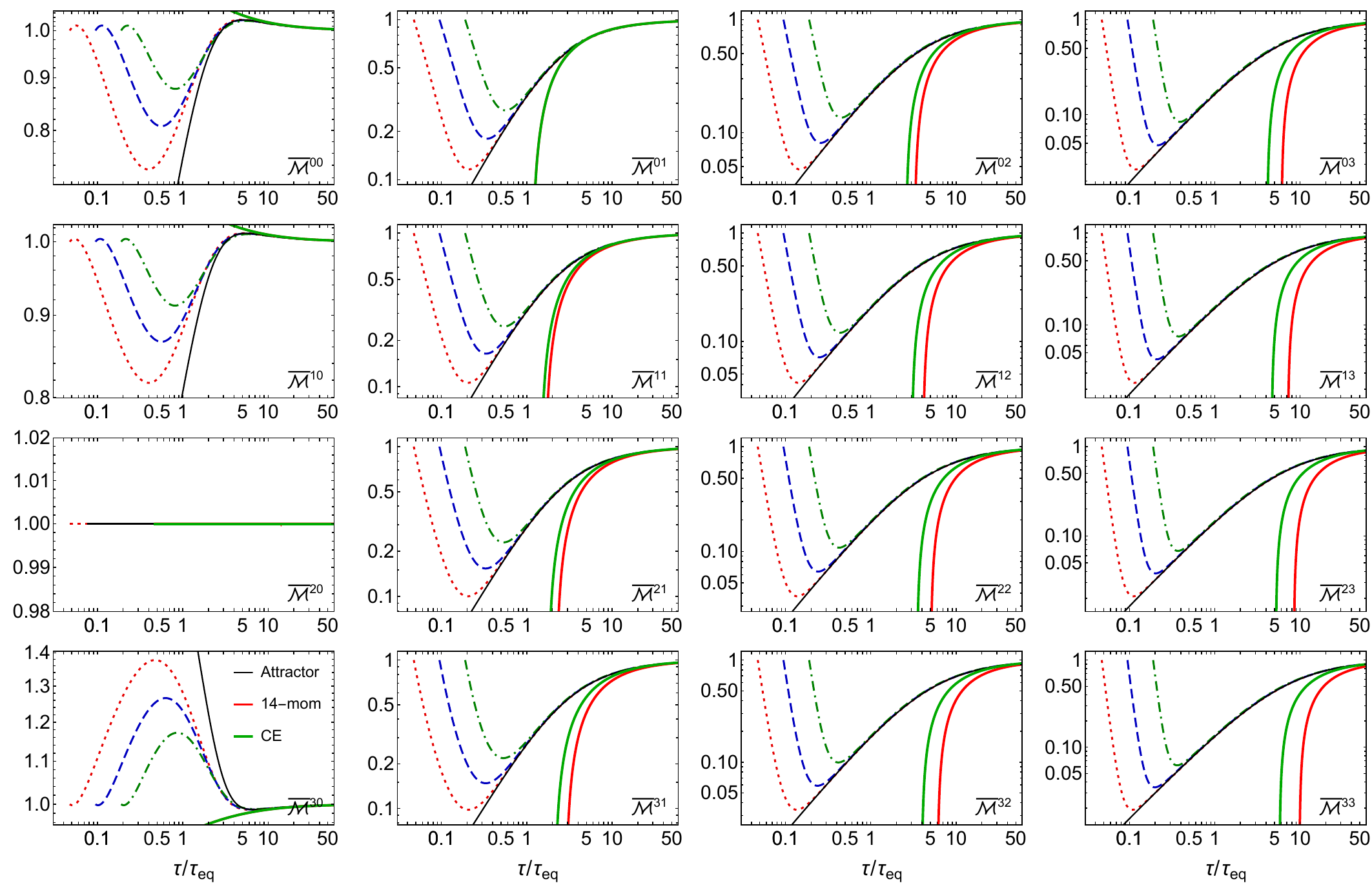}\hspace{2mm}}
\caption{Scaled moments $\Mbar^{nl}$ as a function of rescaled time for the case $m = 1$~GeV.  Initial conditions and line styles are the same as in fig.~\ref{fig:t0scan-m0p2}.}
\label{fig:t0scan-m1}
\end{figure}

\begin{figure}[ht]
\centerline{\includegraphics[width=1\linewidth]{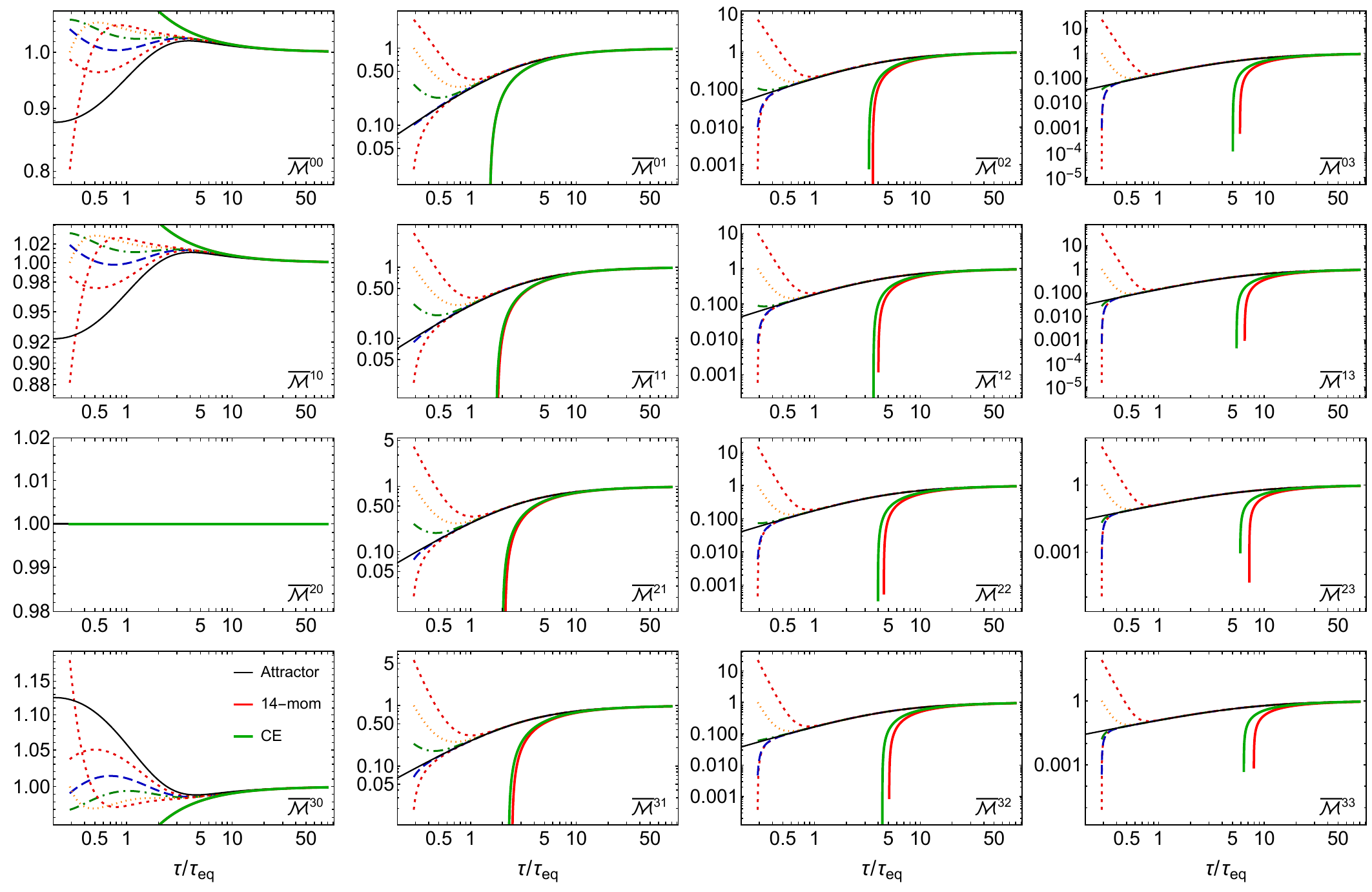}\hspace{2mm}}
\caption{Scaled moments $\Mbar^{nl}$ as a function of rescaled time for the case $m = 5$ GeV.  Initial conditions and line styles are the same as in fig.~\ref{fig:a0scan-m0p2}.}
\label{fig:a0scan-m5}
\end{figure}

\begin{figure}[ht]
\centerline{\includegraphics[width=1\linewidth]{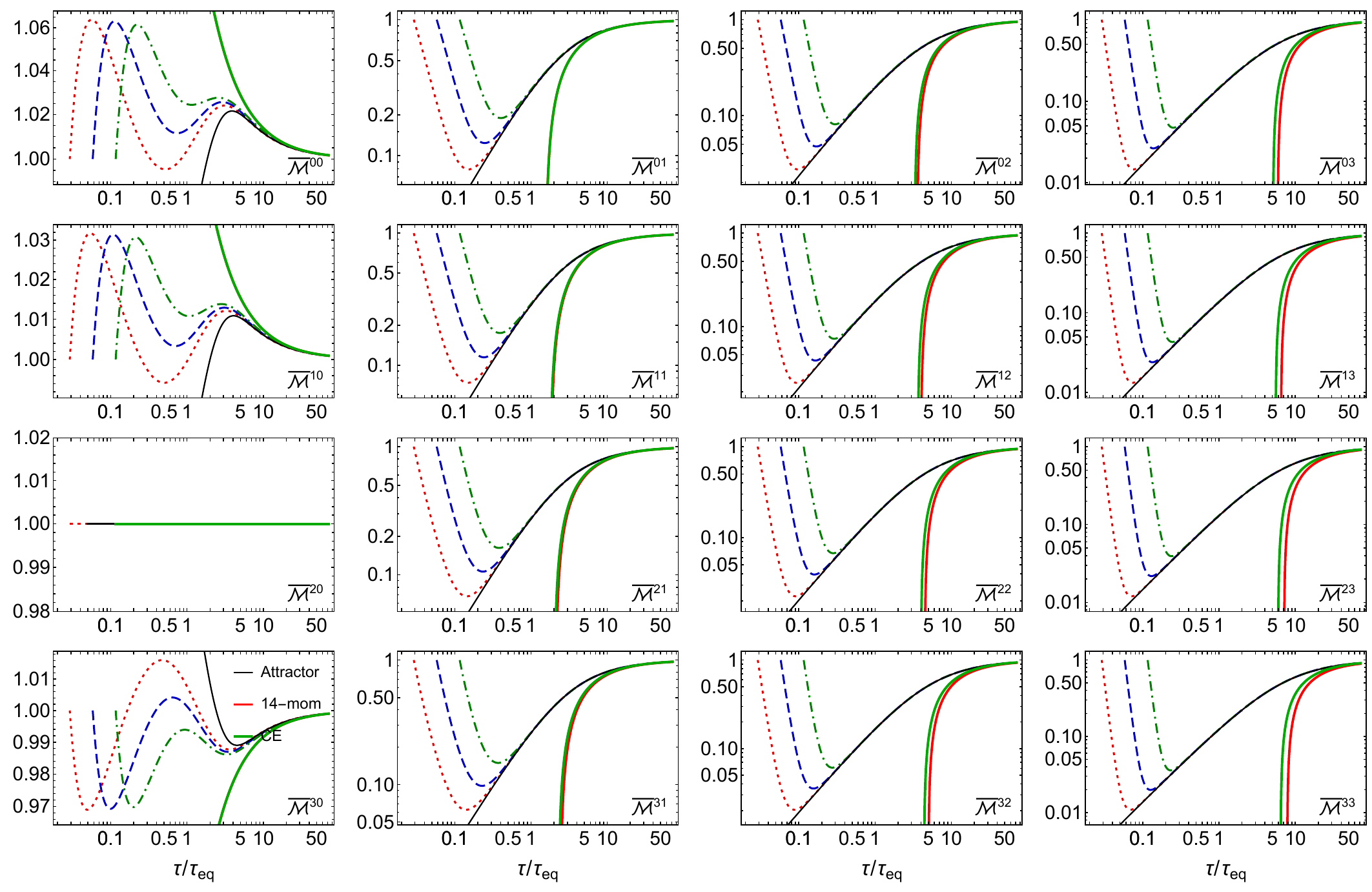}\hspace{2mm}}
\caption{Scaled moments $\Mbar^{nl}$ as a function of rescaled time for the case $m = 5$~GeV.  Initial conditions and line styles are the same as in fig.~\ref{fig:t0scan-m0p2}.}
\label{fig:t0scan-m5}
\end{figure}

\begin{figure}[ht]
\centerline{
\includegraphics[width=0.325\linewidth]{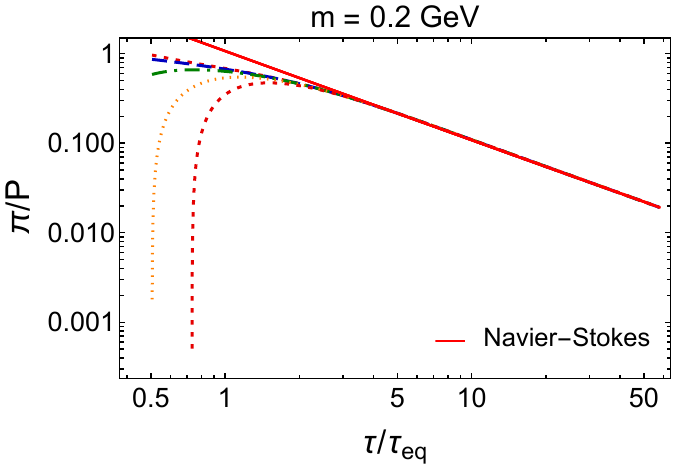}
\includegraphics[width=0.325\linewidth]{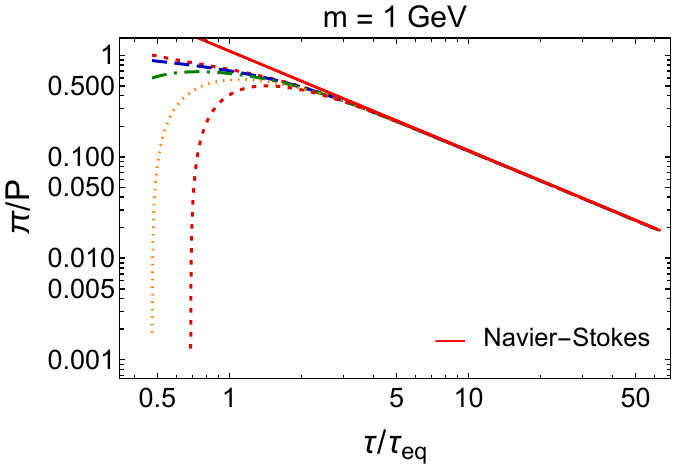}
\includegraphics[width=0.325\linewidth]{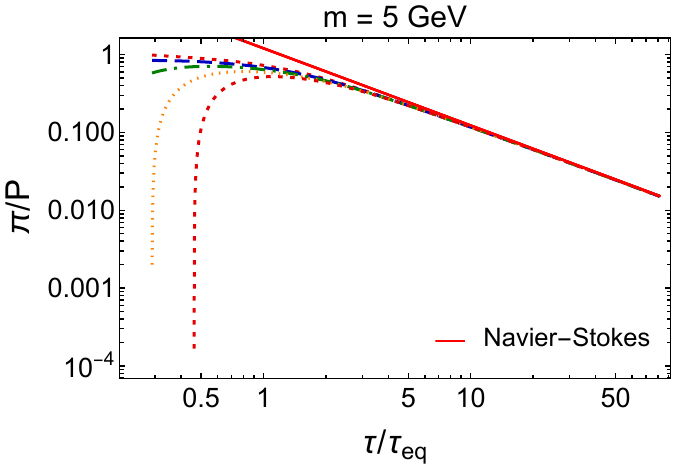}\hspace{2mm}}
\centerline{
\includegraphics[width=0.325\linewidth]{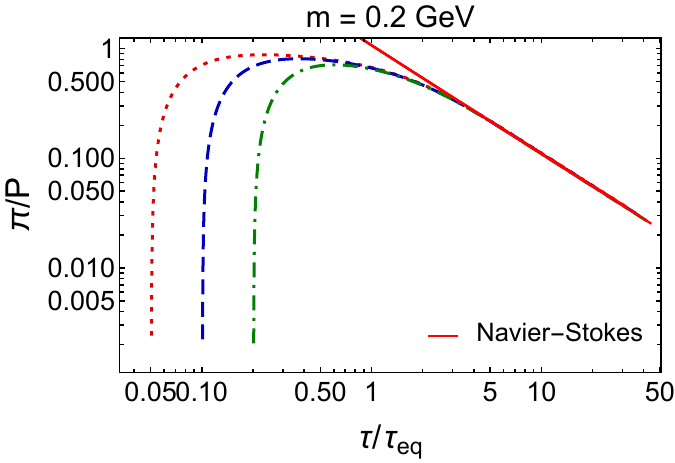}
\includegraphics[width=0.325\linewidth]{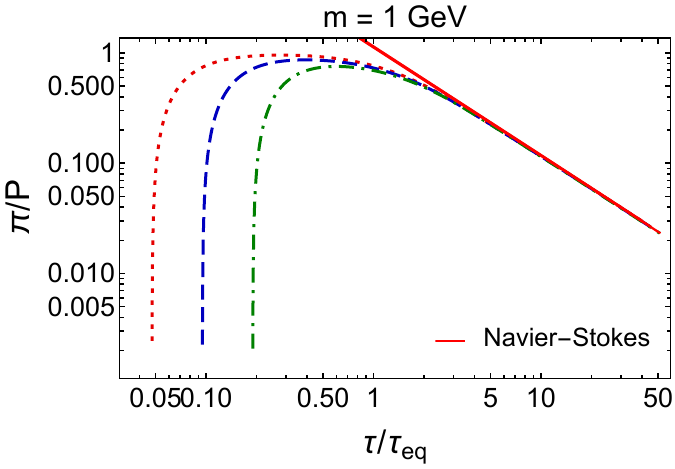}
\includegraphics[width=0.325\linewidth]{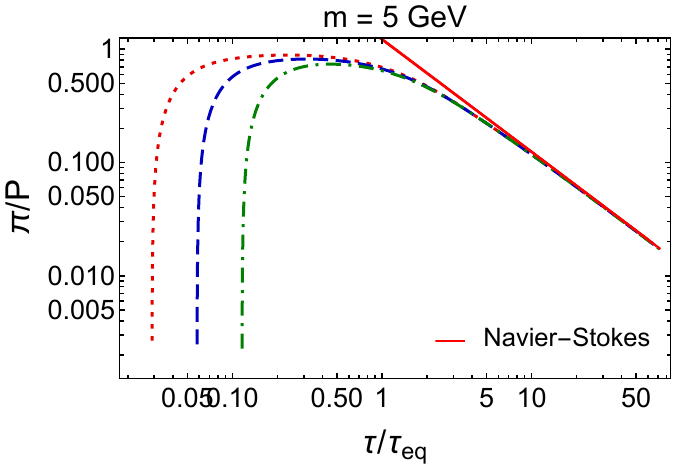}
}
\caption{Scaled shear viscous correction $\pi/P$ from eq.~\eqref{eq:tildeshear} as a function of rescaled time $\tau/\tau_{\rm eq}$.  The top row shows the result of varying the initial anisotropy and the bottom row shows the result of varying the initialization time.  These correspond to the same initializations shown in figs.~\ref{fig:a0scan-m0p2} - \ref{fig:t0scan-m5}.  Columns from left to right show the cases of $m = $ 0.2, 1, and 5 GeV, respectively.  The non-solid curves are specific runs and the solid curve shows the first-order Navier-Stokes prediction given in eq.~\eqref{eq:shearns}.}
\label{fig:shear}
\end{figure}

\begin{figure}[ht]
\centerline{
\includegraphics[width=0.325\linewidth]{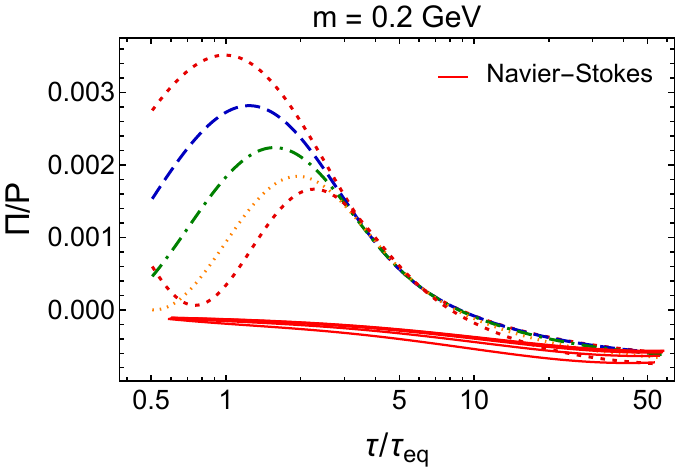}
\includegraphics[width=0.325\linewidth]{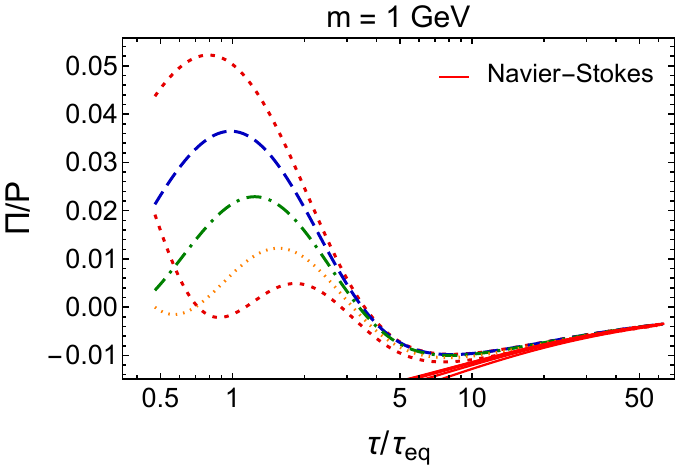}
\includegraphics[width=0.325\linewidth]{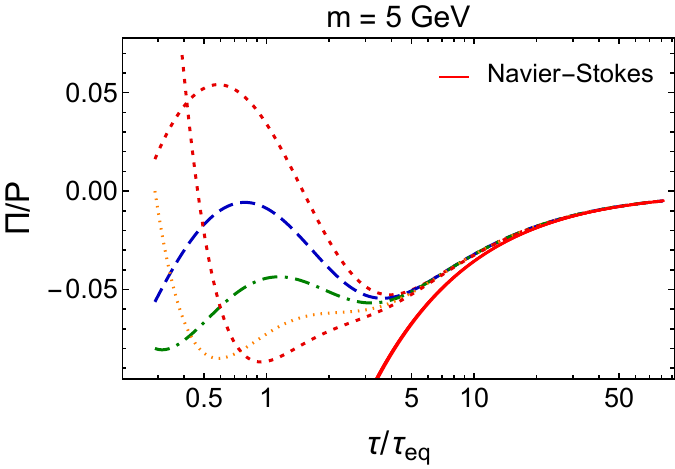}
}
\centerline{
\includegraphics[width=0.325\linewidth]{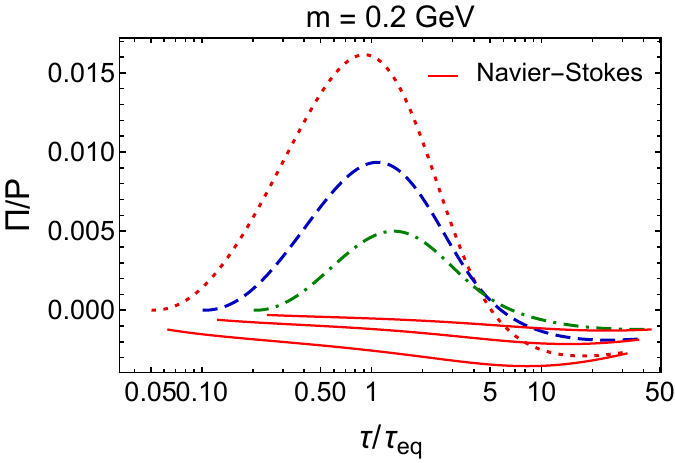}
\includegraphics[width=0.325\linewidth]{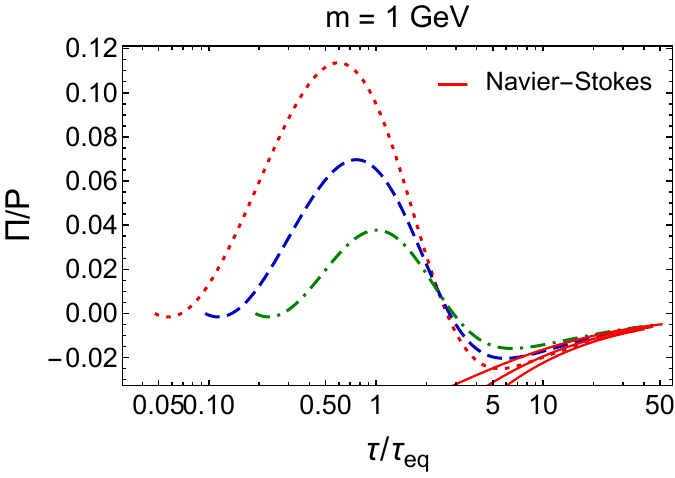}
\includegraphics[width=0.325\linewidth]{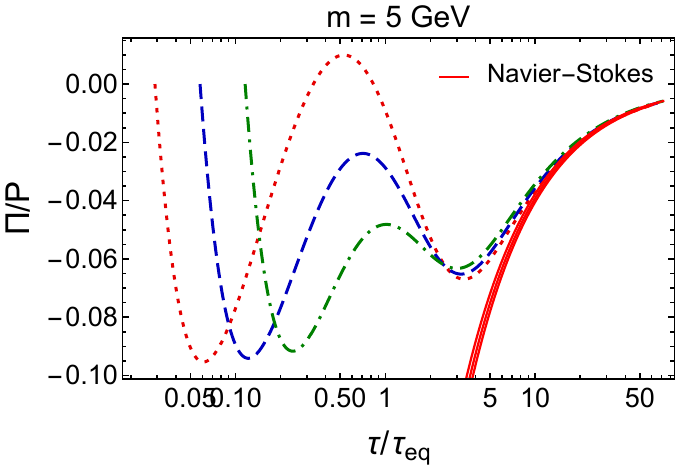}
}
\caption{Scaled bulk viscous correction $\Pi/P$ from eq.~\eqref{eq:tildebulk} as a function of rescaled time $\tau/\tau_{\rm eq}$.  The rows and columns are the same as in fig.~\ref{fig:shear}.  The non-solid curves are specific runs and the solid curve shows the first-order Navier-Stokes prediction given in eq.~\eqref{eq:bulkns}.}
\label{fig:bulk}
\end{figure}

\begin{figure}[ht]
\centerline{
\includegraphics[width=0.325\linewidth]{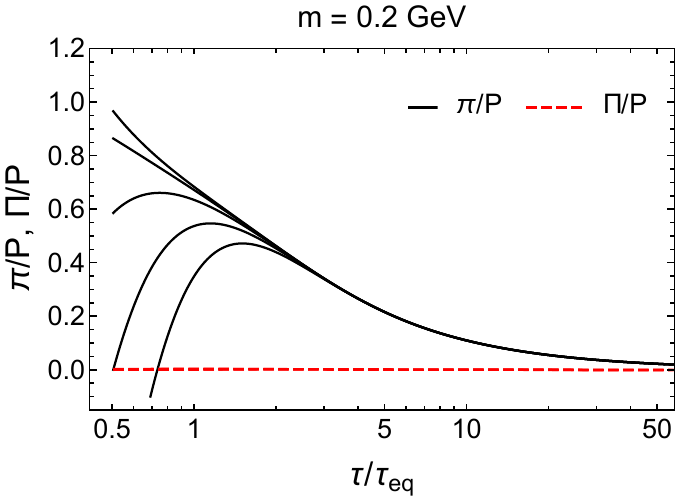}
\includegraphics[width=0.325\linewidth]{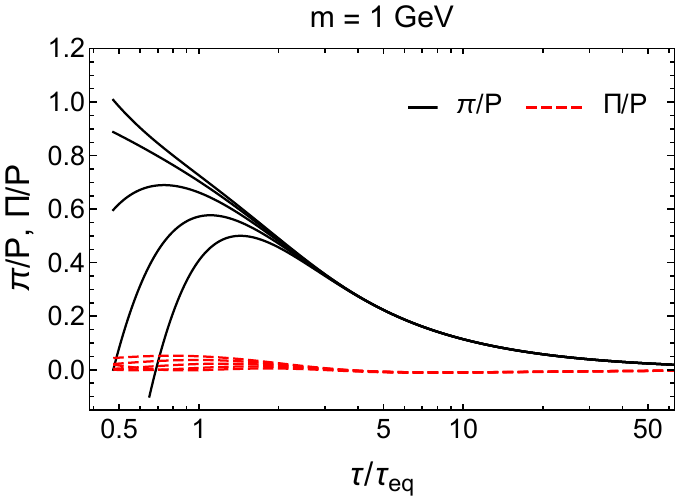}
\includegraphics[width=0.325\linewidth]{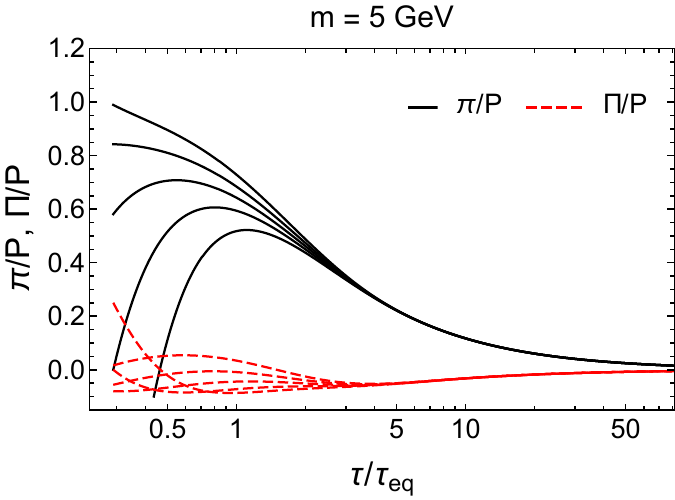}
}
\centerline{
\includegraphics[width=0.325\linewidth]{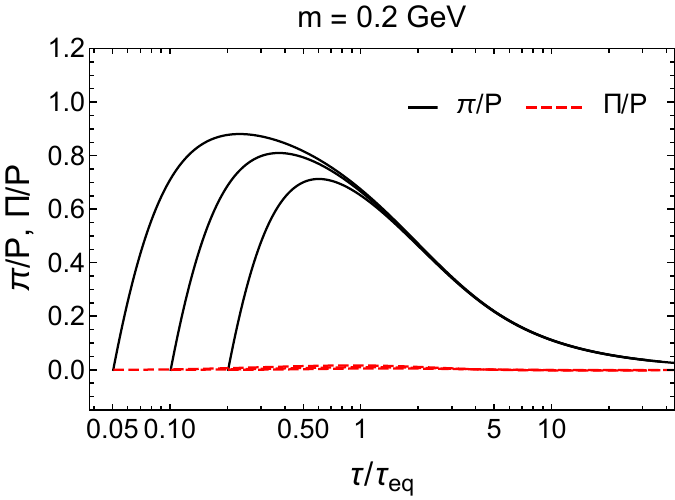}
\includegraphics[width=0.325\linewidth]{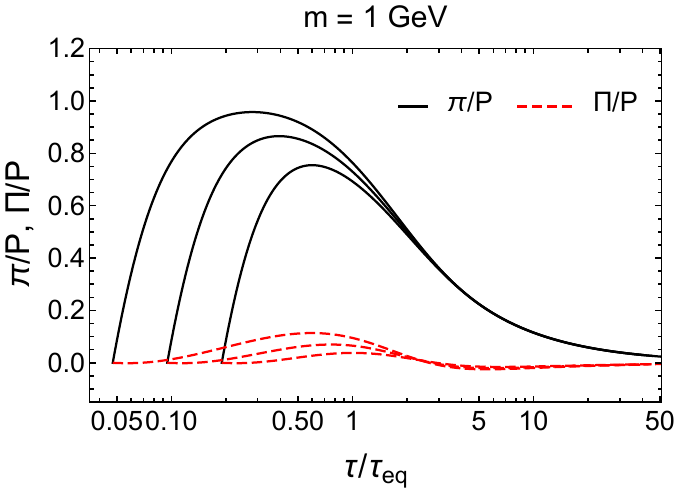}
\includegraphics[width=0.325\linewidth]{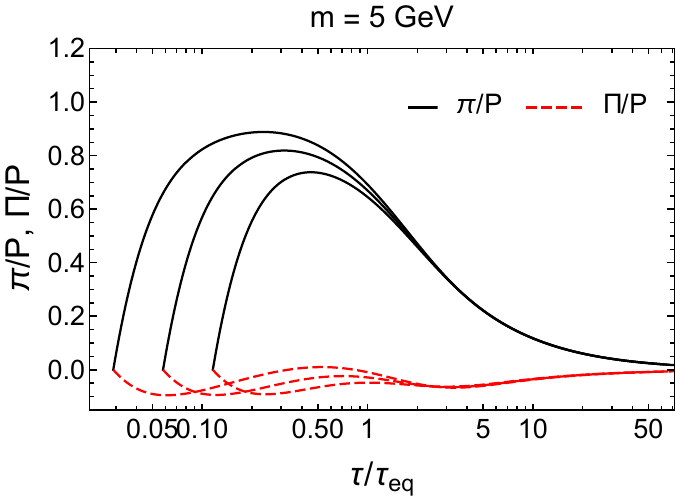}
}
\caption{Scaled shear and bulk viscous corrections from eqs.~\eqref{eq:tildeshear} and \eqref{eq:tildebulk}, respectively, as a function of rescaled time $\tau/\tau_{\rm eq}$.  The black solid lines are the scaled shear correction. and the red dashed lines are the scaled bulk correction.  The rows and columns are the same as in fig.~\ref{fig:shear}.}
\label{fig:shearbulk}
\end{figure}

\subsection{General moments}

In fig.~\ref{fig:a0scan-m0p2} we present our first results for the scaled moments as a function of rescaled time $\wbar = \tau/\tau_{\rm eq}$, which correspond to varying the initial anisotropy while holding the initialization time and initial temperature fixed using a constant mass of $m=0.2$ GeV.  The relaxation time used depends on both the mass and temperature as detailed in eq.~\eqref{eq:teq2}.  The non-solid lines are specific initial conditions initialized at $\tau_0 = 0.1$~fm/c with $T_0=1$~GeV and $\alpha_0 = 1/\sqrt{1+\xi_0} \in \{0.12,0.25,0.5,1,2\}$.  The solid black lines correspond to the attractor solution, the solid red lines are the first-order 14-moment predictions in eq.~\eqref{eq:14momfinal}, and the solid green lines are the first-order Chapman-Enskog predictions in eq.~\eqref{eq:cefinal}.  To obtain the two first-order curves, we evaluated eqs.~\eqref{eq:14momfinal} and eq.~\eqref{eq:cefinal} using the temperature evolution obtained from the exact solution. The attractor lines (black solid line) were obtained by initializing the system at $\tau_0 = 0.01$~fm/c with a high-degree of momentum anisotropy of $\alpha_0 = 1/\sqrt{1+\xi_0} = 0.1$, corresponding to $\xi_0 = 99$.  We note that in this figure and similar panel figures that follow, the fact that the scaled moment $\Mbar^{20}$ is equal to one at all times is due to energy conservation and any deviations from one allow us to gauge the suitability of the discretization used and the convergence of the iterative solution.

As can be seen from fig.~\ref{fig:a0scan-m0p2}, all moments collapse towards the first-order viscous hydrodynamics predictions at late times, with lower-order moments typically converging more quickly than higher-order moments.  For the case of moments with $l=0$, the two first-order schemes Chapman-Enskog and 14-moment coincide identically.  For moments with $l \geq 1$, we find that, for $m=0.2$~GeV, the first-order Chapman-Enskog approximation form for the one-particle distribution function performs better than the 14-moment approximation, particularly for high-order moments.  That said, it is important to emphasize that both fail at early times, with the time scale for breakdown of each scheme becoming larger for higher-order moments.  As demonstrated in ref.~\cite{Strickland:2018ayk}, this continues to be the case if one includes second-order viscous corrections, with only resummed dissipative schemes such as anisotropic hydrodynamics \cite{Florkowski:2010cf,Martinez:2010sc,Tinti:2013vba,Alqahtani:2017mhy,Alalawi:2020zbx,Alalawi:2021jwn} being able to more reliably describe the early-time features of all scaled moments (see in particular the improved schemes introduced in refs.~\cite{Alalawi:2020zbx,Jaiswal:2021uvv}).

Considering earlier times, in fig.~\ref{fig:a0scan-m0p2} we see that for all moments with $l\geq1$ there are indications of a non-equilibrium attractor that extends to very early times.  For the moments with $l=0$, however, we observe that, although the solutions tend towards the attractor, the approach appears to be slower and there doesn't seem to be a complete collapse of the solutions as seen for $l\geq1$.  Finally, we note that for higher-order moments, we see a very rapid collapse to their respective attractors, indicating that the high-momentum region of the one-particle distribution function quickly approaches a universal form.  This is very similar to what occurred in the conformal case \cite{Strickland:2018ayk,Alalawi:2020zbx}.  In those works it was noted that the reason for the slow hydrodynamization of the $l=0$ moments was due to a two-component form of the exact one-particle distribution function, which includes free-streaming and thermalizing components, with the former being highly squeezed along the $p_z$ axis but eventually decreasing in amplitude to a point that it becomes negligible.

In order to better understand whether an early-time attractor exists in this case, in fig.~\ref{fig:t0scan-m0p2} we present the case of holding the initial anisotropy and temperature fixed while varying the initialization time for, again, $m=0.2$~GeV.  In this figure, the non-solid lines are specific initial conditions initialized with $T_0 = 1$~GeV and $\xi_0=0$ at $\tau_0 \in \{0.01,0.02,0.04\}$~fm/c.  The other line styles are the same as in fig.~\ref{fig:a0scan-m0p2}.  As can be see from figure, there clearly exists an early-time attractor for all moments with $l\geq1$.  In the moments with $l=0$ we see a slower approach to the attractor solution, however, the three specific solutions shown approach a semi-universal result at fairly early times in the evolution. For moments with $l\geq1$ we observe that the rate of approach of all specific solutions to the attractor are the same, being associated with the free-streaming period of the evolution.

Turning next to figs.~\ref{fig:a0scan-m1} and \ref{fig:t0scan-m1} we present the result of varying the initial anisotropy and initialization times in the case that $m=1$~GeV.  As before, the initial temperature is held fixed at $T_0=1$~GeV meaning that, in this case, the temperature is always less than or equal to the mass at all times.  From these figures, we see again that there exists both a forward attractor and a pull-back attractor for moments with $l \geq 1$ and that moments with $l=0$ have a slower approach to their respective attractors.  Despite this, the results are still semi-universal after a short time.  With respect to the first-order hydrodynamical forms, we once again see that for higher-order moments, the Chapman-Enskog form provides a more quantitatively reliable asymptotic result than the 14-moment approximation for the higher-order moments, however, both first-order results break down at early times as was seen previously.

Finally, in figs.~\ref{fig:a0scan-m5} and \ref{fig:t0scan-m5} we present the result of varying the initial anisotropy and initialization times in the case that $m=5$~GeV.  Once again, the initial temperature is held fixed at $T_0=1$~GeV.  For this case the temperature is always small compared to the mass scale.  From these figures, we see again that there exists both a forward attractor and a pull-back attractor for $l \geq 1$.  Similar to the other cases, we find that moments with $l=0$ don't seem to possess early-time attractors and, based on fig.~\ref{fig:t0scan-m5}, we see that not even a partial collapse of the different initialization times occurs until around $\tau/\tau_{\rm eq} \sim 3$.  However, we still see a rapid collapse to an attractor for all moments with $l\geq1$.  In particular, we call attention to the panel showing $\Mbar^{01}$, which is equal to the ratio of the longitudinal pressure, $P_L$, divided by the equilibrium longitudinal pressure, $P_{{\rm eq},L} = P$.  A similar collapse of $P_L/P$ was reported in refs.~\cite{Chattopadhyay:2021ive,Jaiswal:2021uvv}, where a conformal relaxation time was employed and a smaller mass of $m = 0.2$ GeV was considered.  Here we have considered even larger masses of 1 and 5 GeV and reached the same conclusion, namely that there is an attractor for $P_L/P$ and we have extended this conclusion to include all moments with $l\geq1$.

\subsection{Bulk and shear viscous corrections}

We now turn to extractions of the shear and bulk viscous corrections from the general moments using eqs.~\eqref{eq:tildeshear} and \eqref{eq:tildebulk}, respectively.  In fig.~\ref{fig:shear}, we present the scaled shear viscous correction $\pi/P$ as a function of rescaled time $\tau/\tau_{\rm eq}$.  The top row shows the result of varying the initial anisotropy and the bottom row shows the result of varying the initialization time.  These correspond to the same initializations shown in figs.~\ref{fig:a0scan-m0p2} - \ref{fig:t0scan-m5}.  Columns from left to right show the cases of $m = $ 0.2, 1, and 5 GeV, respectively.  The non-solid curves are specific runs and the solid curve shows the first-order Navier-Stokes prediction given in eq.~\eqref{eq:shearns}.  As these figures demonstrate, as the mass is increased, there no longer exists an early time collapse of the solutions to a unique attractor curve and the solutions only fully collapse once one enters the region describable by first-order viscous hydrodynamics.

We turn next to fig.~\ref{fig:bulk} where we present the scaled bulk viscous correction $\Pi/P$ \eqref{eq:tildebulk} as a function of rescaled time $\tau/\tau_{\rm eq}$.  The rows and columns are the same as in fig.~\ref{fig:shear}.  
The non-solid curves are specific runs and the solid curve shows the first-order Navier-Stokes prediction given in eq.~\eqref{eq:bulkns}.  
As can be seen from the top row of this figure, only for the smallest mass shown of 0.2 GeV do we see a semi-universal result at early times in the top row and, for the largest mass of 5 GeV, we only see signs of a collapse to a semi-universal curve just prior to the onset of the applicability of first-order hydrodynamics.  
The bottom row of this figure shows that if one reduces the initialization time while holding the initial temperature fixed, there is no early-time attractor and, on top of that, even the late time Navier-Stokes result is not unique.  This should be contrasted with fig.~\ref{fig:shear} where one sees that the late-time Navier-Stokes results for the non-conformal shear collapse to a single line.  
Interestingly, as can be seen from the bottom row of fig.~\ref{fig:bulk}, as the mass is increased, the late-time Navier-Stokes curves begin to collapse to a unique curve, however, there is no indication of a unique early-time attractor.  

In order to put the results for the scaled shear and scaled bulk corrections in a more easily comparable form, in fig.~\ref{fig:shearbulk} we present both in the same panels so that the magnitude of the non-universal behavior can be visualized.  In this figure, the black solid lines are the scaled shear correction. and the red dashed lines are the scaled bulk correction.  The rows and columns are the same as in fig.~\ref{fig:shear}.  As the bottom right panel, in particular, demonstrates, there is a non-trivial cancellation between the shear and the bulk corrections even when the magnitude of the bulk correction is relatively large.

Finally, as was observed in refs.~\cite{Chattopadhyay:2021ive,Jaiswal:2021uvv}, a cancellation of the non-universal features of the scaled shear and bulk corrections occurs, resulting in a universal attractor for the scaled longitudinal pressure, $P_L/P = {\cal \Mbar}^{01}$ which can be clearly see in figs.~\ref{fig:a0scan-m0p2} - \ref{fig:t0scan-m5}.  Such a cancellation occurs in all moments with $l\geq1$ as these figures demonstrate implying that the high-momentum part of the distribution quickly approaches a universal form.  This observation is once again in accordance with the finding of refs.~\cite{Chattopadhyay:2021ive,Jaiswal:2021uvv}, where they presented plots of the scaled-time evolution of the full one-particle distribution function.

\section{Conclusions}
\label{sect:conclusions}

In this paper we have confirmed and extended prior works that studied whether or not attractors exist in non-conformal kinetic theory.  We did this by making use of an exact solution of the boost-invariant Boltzmann equation in relaxation time approximation.  This exact solution is expressed in terms of an integral equation that can be solved numerically by the method of iteration and we derived an integral expression for general moments that allowed us to obtain their time evolution after having solved for the time evolution of the system's temperature.  Associated with this paper we have released the code used for our studies as a publicly available package~\cite{MikeCodeDB}.  Using this method, we studied the time evolution of a large set of integral moments of the one-particle distribution function, varying both the initial momentum-space anisotropy and initialization time, while holding the initial energy density fixed.  We considered three different values of the mass and our main results are presented in figs.~\ref{fig:a0scan-m0p2} - \ref{fig:t0scan-m5}.  From the time evolution of the general moments, we were able to compute the exact time evolution of both the shear and bulk viscous corrections to the one-particle distribution function and we presented these in figs.~\ref{fig:shear} - \ref{fig:shearbulk}, where we compared them to their corresponding expressions at leading-order in the gradient expansion.

Our conclusions from this study are consistent with those found by the authors of refs.~\cite{Chattopadhyay:2021ive,Jaiswal:2021uvv}, namely that there exists both late- and early-time attractors for the scaled longitudinal pressure $P_L/P$, while these do not exist separately for the shear and bulk viscous corrections.  In terms of the moments, this is manifested in the fact that moments with $l=0$ for the $n$ values considered herein ($n\leq3$), do not seem to possess an early-time attractor.  In order to assess the approach to the late-time hydrodynamic attractor, we derived expressions for the viscosity-corrected one-particle distribution function to leading order in the gradient expansion (Navier-Stokes limit) within both the 14-moment and Chapman-Enskog approximations.  We found that, for small masses, the first-order Chapman-Enskog form was quantitatively more reliable at late times than the 14-moment approximation, particularly for higher-order moments; however, for larger masses, the two approximations resulted in quantitatively similar results when compared to the exact solutions.  Finally, in order to connect to standard viscous hydrodynamics corrections, we extracted the time evolution of both the shear and bulk viscous corrections from the exact solution.  One new observation on this front is contained in fig.~\ref{fig:bulk}, where it can be seen that the bulk viscous correction at first-order in gradients does not collapse in the late-time, Navier-Stokes, limit.

As to the practical implications of our results we note that, in the conformal case ref.~\cite{Strickland:2017kux} demonstrated that even the $l=0$ moments possessed a universal forward attractor and that this implied that there was an attractor for the full one-particle distribution function.  In the non-conformal case, the authors of Ref.~\cite{Jaiswal:2021uvv} presented results for the full one-particle distribution function, finding that apart from slower convergence to the attractor at very low longitudinal momentum, it exhibited attractor behavior as well.  Although their study was restricted to a conformal relaxation time, our work indicates that the same conclusion would be reached with a non-conformal relaxation time. 

Finally, as to the implications for heavy-ion phenomenology, it has been shown that in the conformal case the existence of a longitudinal pressure attractor can be used, e.g. to constrain the initial energy density of the QGP \cite{Giacalone:2019ldn} and electromagnetic emissions \cite{Coquet:2021lca}.  Since the arguments therein only rely on their being a longitudinal pressure attractor, it seems that they would go through unchanged.  Our results, at worst, indicate that may be some additional uncertainty associated with such treatments if, in the future, they were to rely on attractors existing also in the case $l=0$.  Because of this, the overarching idea to use attractors in this manner would still be sound.  This is due to the fact that, when considering the forward attractor with phenomenologically relevant initialization times, e.g. \mbox{0.1 fm/c}, we still see a universal collapse to the forward attractor for all moments with $l \neq 0$ and a semi-universal collapse for the moments with $l = 0$ (see e.g. figs.~\ref{fig:a0scan-m0p2}).  This semi-universality would introduce a small degree of uncertainty in the conclusions compared to the conformal case, but would not make this a useless exercise.

Looking to the future, it would be very interesting to see if the conclusions contained herein can be extended to the case of a quasiparticle Boltzmann gas with temperature-dependent masses.  Such a picture underpins quasiparticle anisotropic hydrodynamics and allows it to make use of a realistic non-conformal equation of state~\cite{Alqahtani:2015qja,Alqahtani:2017jwl,Alqahtani:2017tnq,Almaalol:2018gjh,Alqahtani:2020paa}.  It will also be interesting to see if these findings are modified if one includes the effect of dynamical 2+1D and 3+1D expansion \cite{Romatschke:2017acs} and thermal noise \cite{Chen:2022ryi}.  We leave these considerations to future works.

\acknowledgments

We thank Chandrodoy Chattopadhyay and Ulrich Heinz for discussions. H.A. was supported by the Deanship of Scientific Research at Umm Al-Qura University under Grant Code 22UQU4331035DSR01.  M.S. was supported by the U.S. Department of Energy, Office of Science, Office of Nuclear Physics under Award No. DE-SC0013470.

\appendix

\section{Generalized initial condition}
\label{app:a}

\begin{figure}[th]
\centerline{\includegraphics[width=1\linewidth]{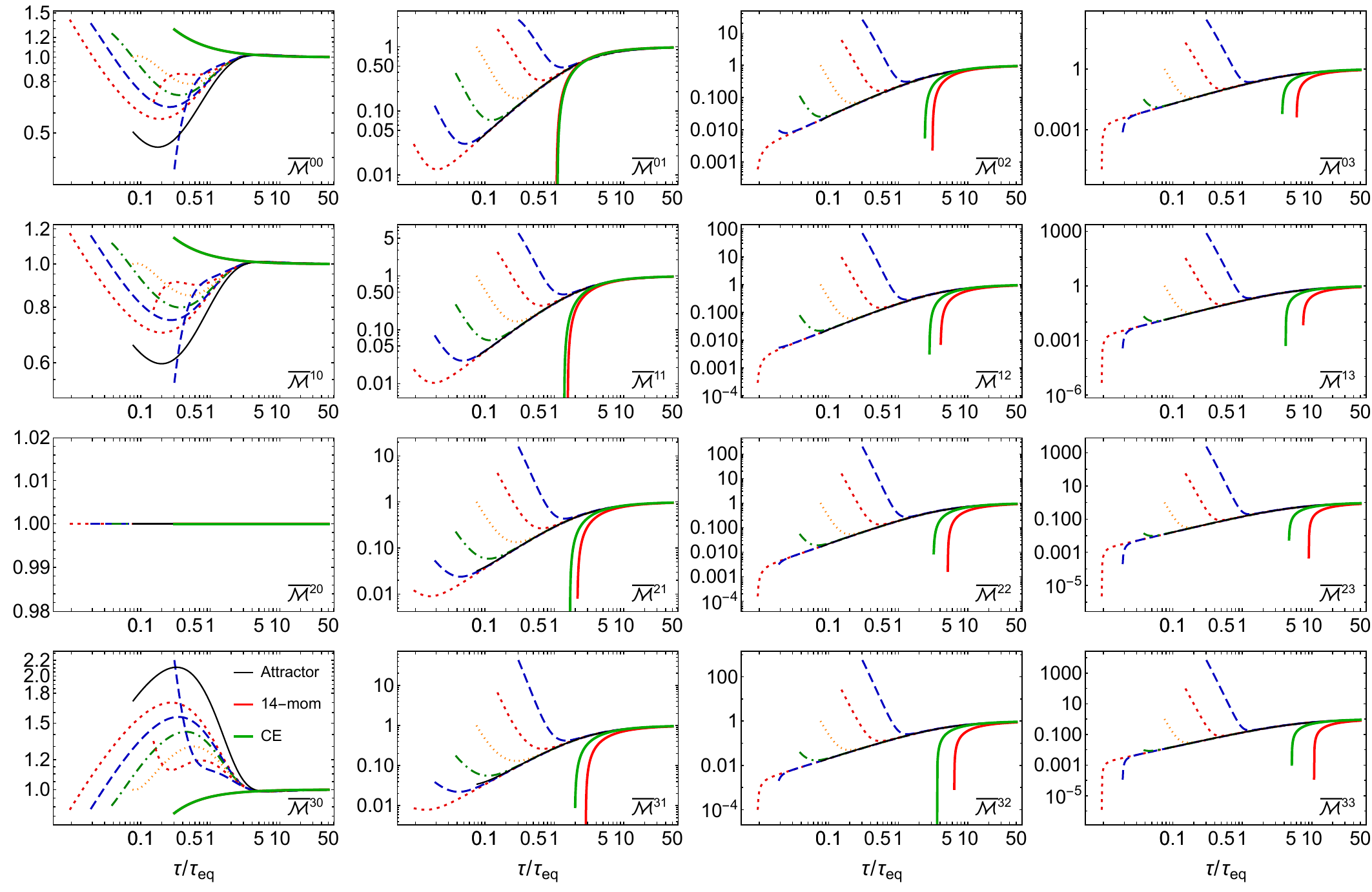}\hspace{3mm}}
\vspace{-3mm}
\caption{Scaled moments $\Mbar^{nl}$ as a function of rescaled time for the case $m = 1$~GeV obtained using the generalized spheroidal initial condition specified in eq.~\eqref{eq:jaiswal} and varying all parameters appearing therein, $\xi_0$, $\gamma_0$, and $\tau_0$.}
\label{fig:allscan-m1}
\end{figure}

\begin{figure}[th]
\centerline{
\includegraphics[width=0.325\linewidth]{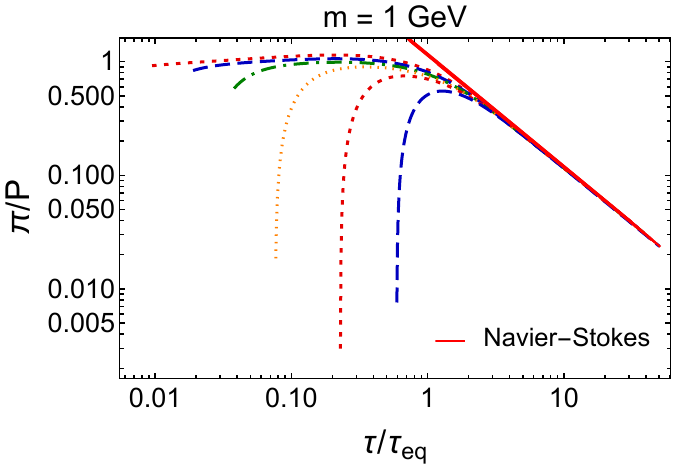}
\includegraphics[width=0.325\linewidth]{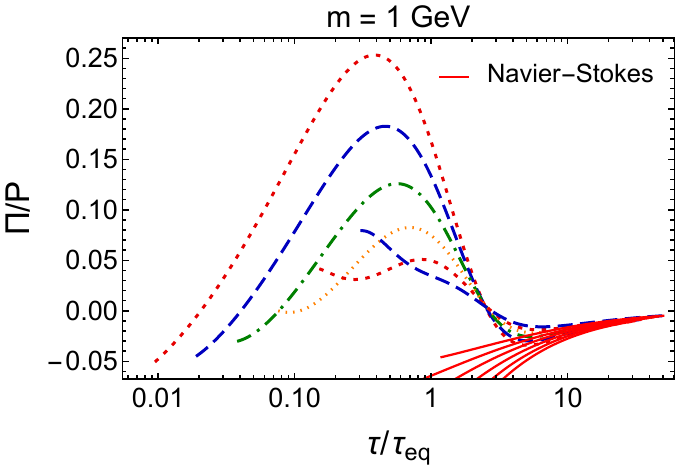}
\includegraphics[width=0.325\linewidth]{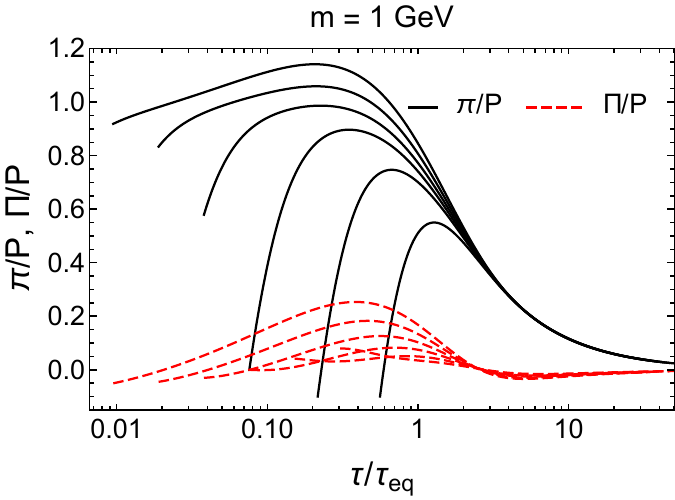}
}
\vspace{-3mm}
\caption{Scaled shear viscous correction $\pi/P$ (left), bulk viscous correction $\Pi/P$ (middle), and both combined (right) as a function of rescaled time $\tau/\tau_{\rm eq}$ for $m=1$ GeV and varying all parameters appearing in eq.~\eqref{eq:jaiswal}, $\xi_0$, $\gamma_0$, and $\tau_0$, while holding the initial energy density fixed to that of an isotropic equilibrium gas with $T_0 = 1$ GeV.}
\label{fig:bulkshear-allscan-m1}
\end{figure}

In this appendix we present results obtained using the generalized spheroidal initial condition introduced in ref.~\cite{Jaiswal:2021uvv}
\begin{equation}
f_0(w,p_T) = \frac{1}{\gamma_0}
\exp\left[
-\frac{\sqrt{(p\cdot u)^2 + \xi_0 (p\cdot z)^2}}{\Lambda_0} \, \right] ,
\label{eq:jaiswal}
\end{equation}
where $\gamma_0$ allows us to independently vary the initial shear and bulk corrections.  Since there are now three independent parameters to vary, we consider here varying all of them simultaneously, while holding the initial energy density fixed to that of an isotropic equilibrium gas with $m=1$ GeV and $T_0 = 1$ GeV.  In figs.~\ref{fig:allscan-m1} and \ref{fig:bulkshear-allscan-m1} we present the evolution of the scaled moments and viscous corrections resulting from such a scan.  As these figures demonstrate, as with the spheroidal initial conditions used in the main body of the text, there does not seem to be a pull-back attractor for moments with $l=0$ nor the viscous corrections $\pi$ and $\Pi$, while both forward and pull-back attractors are still evident for all moments with $l \neq 0$.

\bibliographystyle{JHEP}
\bibliography{attractor}

\end{document}